\documentclass[reprint,
superscriptaddress,
amsmath,
amssymb,
aps,
twocols,
nofootinbib
]{revtex4-1} %REVTEX
\usepackage[T1]{fontenc}
\usepackage[utf8]{inputenc} 
\usepackage[english]{babel}
\usepackage{physics}
\usepackage{amsthm} 
\usepackage{amsfonts}
\usepackage{mathtools}
\usepackage{verbatim}
\usepackage{graphicx}
\usepackage{dcolumn}
\usepackage{bm}
\usepackage[caption=false]{subfig}
\usepackage{hyperref}
\usepackage[usenames, dvipsnames]{color}

\begin{document}
\title{Low-resolution descriptions of model neural activity\\ reveal hidden features and underlying system properties}

\author{Riccardo Aldrigo}
\affiliation{Physics Department, University of Trento, via Sommarive, 14 I-38123 Trento, Italy}
\author{Roberto Menichetti}
\affiliation{Physics Department, University of Trento, via Sommarive, 14 I-38123 Trento, Italy}
\affiliation{INFN-TIFPA, Trento Institute for Fundamental Physics and Applications, I-38123 Trento, Italy}
\author{Raffaello Potestio}
 \email{raffaello.potestio@unitn.it}
\affiliation{Physics Department, University of Trento, via Sommarive, 14 I-38123 Trento, Italy}
\affiliation{INFN-TIFPA, Trento Institute for Fundamental Physics and Applications, I-38123 Trento, Italy}

\date{\today}

\begin{abstract}
The analysis of complex systems such as neural networks is made particularly difficult by the overwhelming number of their interacting components. In the absence of prior knowledge, identifying a small but informative subset of network nodes on which the analysis should focus is a rather challenging task. In this work, we address this problem in the context of a Hopfield model, which is observed through the lenses of low-resolution representations, or decimation mappings, consisting of subgroups of its neurons. The optimal, most informative mappings of the network are defined through a recently developed methodology, the mapping entropy optimisation workflow (MEOW), which performs an unsupervised analysis of the states sampled by the network and identifies those subgroups of spins whose configuration distribution is closest to that of the full, high-resolution model. Which neurons are retained in an optimal mapping is found to critically depend on the properties of the interaction matrix of the network and the level of detail employed to describe the system; by these means, it is thus possible to extract quantitative insight about the underlying properties of the high-resolution model through the analysis of its optimal low-resolution representations. These results show a tight and potentially fruitful relation between the level of detail at which the network is inspected and the type and amount of information that can be gathered from it, and showcase the MEOW approach as a practical, enabling tool for the study of complex systems.
\end{abstract}

\maketitle

\section{Introduction} \label{Sec:introduction}

Neural systems owe their complex emergent phenomena to the winding, nonlinear interplay of a large number of fundamental units -- neurons \cite{rolls1998neural,dayan2005theoretical,zamora2011characterizing,lynn2019physics}. The intricacy and sheer size of biological neural networks (the human brain is made of tens of billions of nerve cells \cite{herculano2009human,joudaki2012eeg}) is however such that a comprehensive understanding of their behaviour is still largely out of reach; several approaches have thus been proposed that aim, \emph{via} the introduction of rather essential and computationally manageable models, at reproducing and/or capturing specific aspects of neural activity. These effective representations of the brain architecture and function range, e.g., from generalised Potts models \cite{kropff2005storage,russo2008free} or restricted Boltzmann machines \cite{ackley1985learning,aarts1989simulated,marullo2020boltzmann,goel2020boltzmann} to graph-theory methods \cite{sporns2014network,bassett2018nature} and the most recent, large-scale deep learning techniques \cite{vogels2005annurev}. 
In this context, a classic example of \textit{in silico} neural network is provided by the Hopfield model \cite{hopfield1982neural,hopfield1984pnas,stat_mech_H_model,ramsauer2020hopfield}, in which a neuron is represented as a two-state variable, or binary spin, whose values are associated with the biological \emph{firing} or \emph{rest} conditions; albeit their relatively simple mathematical formulation, Hopfield models---or extensions thereof---showcase a rich phenomenology, and are still widely employed to investigate the process of memory retrieval \cite{barra2010replica,agliari2018non,agliari2020replica,alemanno2023hopfield}.

Irrespective of their specific details and purpose, the aforementioned models exhibit several advantageous features: on the one hand, the overall ``simplicity'' of their elemental constituents and associated interactions, as well as the controllable size of the network---where the latter can be either sufficiently small to carry out numerical simulations \cite{crisanti1986saturation,bruck1990convergence,stat_mech_H_model,ramsauer2020hopfield}, or infinitely large to allow an exact mathematical treatment \cite{huang2021statistical}; on the other hand, one has that the results of an analysis of the emergent properties of the system can be interpreted in light of the structure of the underlying model, which is known by definition.
Critically, this does not generally hold in the study of biological neural networks: here, in fact, it is rarely---if not outright never---the case that one can acquire data (e.g. time series) about the state of each individual neuron in the system; as a particularly evident example, think of an electroencephalogram (EEG) exam where only a few tens of signal streams are recorded out of the billions of neurons that compose a biological brain \cite{herculano2009human,joudaki2012eeg}. Additionally, the detailed characteristics of the network that are responsible for the \emph{generation} of those states are usually unknown: taking once again the example of an EEG, the neurons that contribute to the experimentally observed patterns are too complex, too many, and their connections too tightly intertwined for all these ingredients to be deconstructed in sufficient detail.

In the framework of an empirical analysis of a subset of neurons, whereby only the emergent behaviour of the system is observed while its generative mechanism is unknown, it would be thus desirable to develop strategies that allow one to distinguish between particularly ``important'' units, whose states reveal relevant information about the system, and ``irrelevant'' ones that can be safely ignored in the study.
In this work, we employ an information-theoretic analysis method recently developed by some of us, namely the mapping entropy optimisation workflow (MEOW) \cite{JCTC, errica2021deep, HGP, giulini2024excogito}, to identify and characterise maximally informative subsets of neurons in a network only given the time series of their states; we thus aim to identify low-resolution representations of the system in terms of small(er) numbers of elements that can be almost as useful as a fine-grained description that accounts for all the network constituents. The MEOW strategy, originally introduced in the context of the analysis of complex biomolecular structures \cite{JCTC,giulini2024excogito}, relies on the idea that a subgroup of elements is ``important'' if observing them provides (almost) as much statistical information as one gathers by analysing the whole system; more specifically, we take the empirical probability distribution of the observed network states as the high-resolution reference, and attempt at reconstructing it from the distribution of low-resolution configurations of a subset of neurons. The discrepancy between the original, high-resolution empirical probability distribution and the reconstructed one is quantified by the \emph{mapping entropy}, a Kullback-Leibler divergence between them \cite{kullback1951information,Shell_Scott,rudzinski2011coarse,noid2023perspective}. The MEOW strategy searches for those subgroups of elements that minimise the mapping entropy, constituting the maximally informative reduced representations of the network that can be designed to investigate its behaviour.

We here apply this idea to the case of a Hopfield model. While the MEOW approach only takes the time series of the neuron states as an input, applying the protocol in such a context enables us to directly relate the outcomes of the analysis to the structure and interactions of the model. Hence, this allows us to validate the workflow in a controlled case where the details of the system are known, focussing on the effect that specific realisations of the memory patters have on the resulting emergent behaviour of the network. This analysis paves the way to the application of MEOW in more complex scenarios in which only the observations of the network states are available, while the underlying generative process is not.

The MEOW protocol enables us to highlight a number of system properties related to the level of detail at which the Hopfield network is observed: in particular, we identify three distinct regimes in which qualitatively different groups of neurons are pinpointed as informative depending on the resolution level of the coarse description, the latter being roughly related to the number of constituents retained in the simplified picture. These results highlight the tight connection between the level of detail at which a system is described and the amount and quality of information that its analysis can reveal.

The paper is organized as follows. In Sect. \ref{sec:matandmeth} we recap the fundamentals about the Hopfield model and the mapping entropy minimisation workflow. In Sect. \ref{sec:results} we illustrate and comment on the results of the MEOW analysis of various types of Hopfield networks of different sizes. Lastly, in Sect. \ref{sec:conclusions} we provide our concluding thoughts and discuss possible future developments and applications of this work.

\section{Materials and Methods}\label{sec:matandmeth}

In this section we provide an overview of the fundamental ingredients of the models and analysis methods employed in this manuscript. More specifically, in Sec.~\ref{subsec:Hopfield_model} we briefly introduce the Hopfield model, summarise its main properties, and discuss the technical aspects of the numerical simulations carried out in this work. Subsequently, in Sec.~\ref{subsec:mapping_entropy} we recap the strategy and the constitutive information-theoretic quantities underlying the mapping entropy optimization workflow recently developed by some of us \cite{JCTC, errica2021deep, HGP, giulini2024excogito}, focusing on those aspects that are specifically associated with the application of MEOW to the analysis of a Hopfield network.

\subsection{The Hopfield model}\label{subsec:Hopfield_model}

The neural network model analysed in this work was originally developed by J.J. Hopfield in 1982 with the aim of exploiting the collective properties of the system as content-addressable memories \cite{hopfield1982neural}. Starting from previous studies in the field \cite{McCulloch, Rosenblatt, Minsky_Papert}, in his seminal manuscript Hopfield formulated the first example of an \emph{attractor} neural network (ANN)~\cite{amit1985prl,MBF}, namely a network capable, in appropriate conditions, of retrieving a set of stored memory patterns encoded in the system's interaction matrix, leveraging a nonlinear dynamic evolution of its constituents.
    
\begin{figure}[t]
    \centering
    \includegraphics[width=0.8\columnwidth]{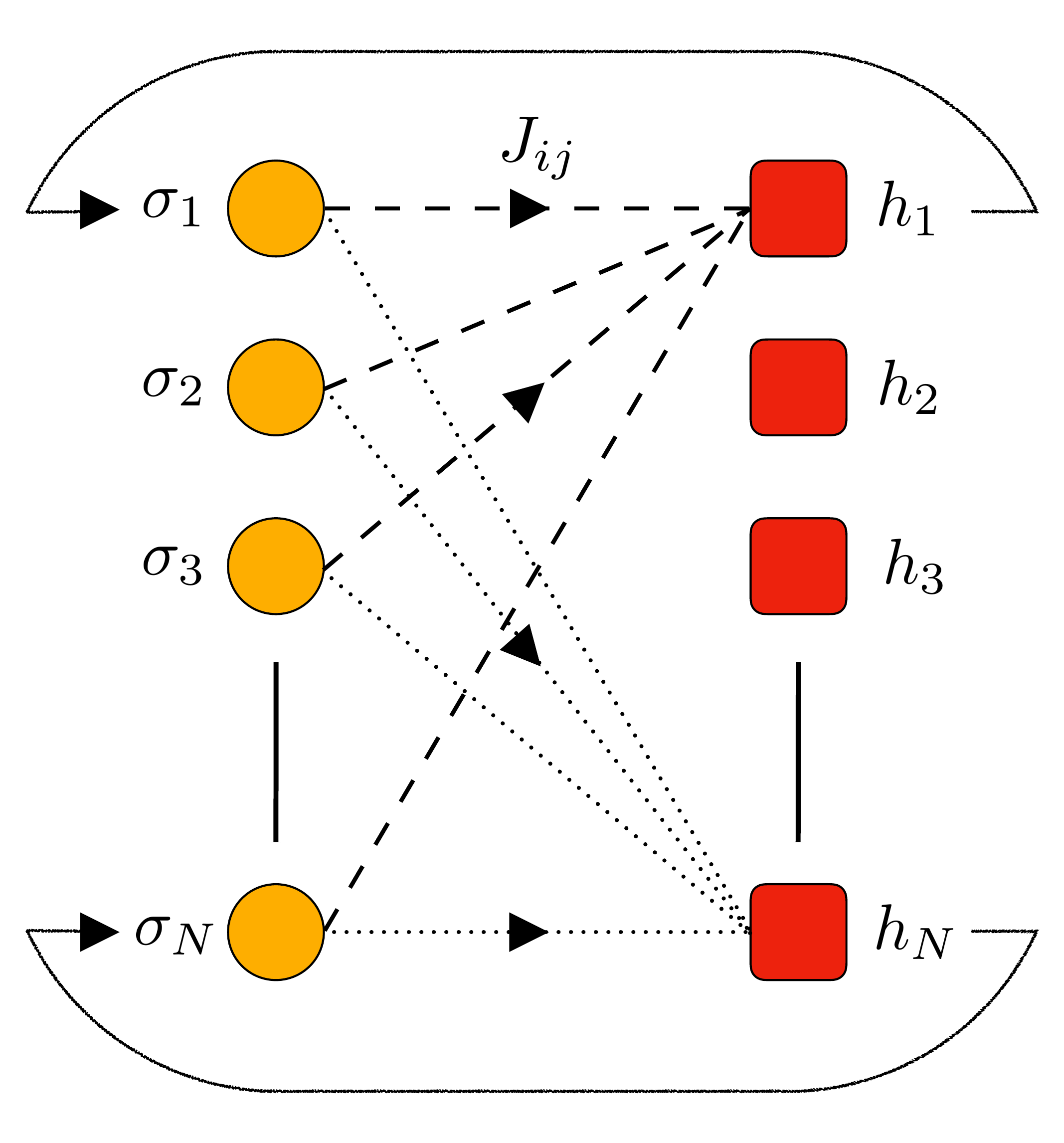}
    \caption{Schematic illustration of the recurrent architecture underlying the dynamics of the Hopfield model. A multi-perceptron system is closed onto itself to form an attractor neural network, with the states $\sigma_i, \ i=1,..., N$ of all neurons at a given time being combined via the synaptic coefficients $J_{ij}$ to provide the set of local fields $h_i$. Such fields, after undergoing a nonlinear transformation, constitute the state of the network at the subsequent timestep.}
    \label{fig:ANN}
\end{figure}

A Hopfield network consists of a single layer of $N$ coupled perceptrons, or model neurons, each of which can be represented as a two-state spin $\sigma_i, \ i=1,..., N$ that can take on $+ 1$ and $-1$ values, respectively associated to the neuron being in a firing (active) and silent (inactive) condition. The system dynamics is based on a recurrent architecture, meaning that the output states of the perceptrons are employed as inputs of the same layer that generates them. More specifically, the state $\{\sigma_i(t)\}$ of the network at time $t$ is processed by the neurons to form the set of output signals $\{h_i(t)\}$; such \emph{local fields} constitute, at the subsequent time cycle and after being subjected to a nonlinear transformation, the new input state $\{\sigma_i(t + \Delta t)\}$ of the network, see Fig.~\ref{fig:ANN} for a schematic representation of this overall workflow. As a consequence of this dynamic evolution, starting from an initial state $\{\sigma_i(t_0)\}$ the system over time wanders throughout the space of the $2^N$ possible configurations available to its neurons, eventually converging towards one of the stored memories \cite{MBF}.

Critically, the memories consist of a set of $p$ states, or \emph{patterns} $\{\xi^\mu_i\}$,
\begin{eqnarray}
\label{eq:hopfield_memories}
    &&\{\xi^\mu_i\} \equiv (\xi_1^\mu, \xi_2^\mu, \cdots\ \xi_N^\mu), \nonumber \\ 
    &&\xi^{\mu}_i = \pm 1, \nonumber \\ 
    &&\mu = 1,...,p.
\end{eqnarray}

A typical statistical-mechanical analysis of the model in the thermodynamic limit relies on the assumption that the memories are independently distributed random variables, with
\begin{equation}
    P(\{\xi^{\mu}_i\})= \prod_{\mu=1}^p\prod_{i=1}^N\left(\frac{1}{2}\delta(\xi^{\mu}_i,1)+\frac{1}{2}\delta(\xi^{\mu}_i,-1) \right),
\end{equation}
where $\delta(\cdot,\cdot)$ represents a Kronecker delta. This, however, is not a necessary requirement, and the memory patterns can indeed entail nontrivial statistical properties.

From the preceding discussion, it follows that the behaviour of a Hopfield network is dictated by three main ingredients, namely (\emph{i}) how the state of the network at time $t$ is processed to return the set of local fields; (\emph{ii}) how the latter relate to the configuration of the system at the following time cycle; and (\emph{iii}) how the memory patterns enter in the definition of the model. Let us briefly analyse these three aspects.

As for the first ingredient, the Hopfield model resorts to a superposition principle for which the total signal received by a specific perceptron is given by a linear combination of those that are transmitted to it by the remaining $N-1$ neurons in the system. The local field $h_i(t)$ experienced at time $t$ by the $i$-th perceptron thus reads
\begin{equation}
    h_i(t) = \sum_{j=1}^N J_{ij} \sigma_j(t),
    \label{eq:PSP_1}
\end{equation}
see Fig.~\ref{fig:ANN}, where the time-independent couplings $J_{ij}$ in Eq.~\ref{eq:PSP_1}, with $J_{ii}=0$, characterise the interaction between each pair of neurons in the system and are called \emph{synaptic coefficients} of the network.

Different prescriptions can then be employed to transform the local fields into the new state of the network, which can be however divided into two main categories depending on whether the nature of the nonlinear relation between $\{h_i(t)\}$ and $\{\sigma_i(t + \Delta t)\}$ is deterministic or stochastic~\cite{MBF}. In this work, we rely on a the stochastic time evolution of the system implemented through a noisy Glauber dynamics \cite{Glauber}, in which the probability $P(\sigma_i(t + \Delta t)=\sigma_i\mid h_i(t)=h_i)=P(\sigma_i\mid h_i)$ that the $i$-th perceptron turns into the firing state $\sigma_i=\pm 1$ when subject to a field $h_i$ reads
\begin{equation}
\label{eq:noisy_dynamics}
    P(\sigma_i\mid h_i)= \frac{\exp(\beta h_i \sigma_i)}{\exp(\beta h_i) + \exp(-\beta h_i)},
\end{equation}
where $\beta^{-1}=T$ is an effective temperature parameter quantifying the influence of the noisy environment on the synaptic transmission. For $T = 0$ the system is driven towards its lowest-energy state(s).

Neurons are evolved asynchronously \cite{Amit_1, Amit_2_saturation, MBF, hopfield1982neural}, so that each time cycle $\Delta t$ consists of a series of $N$ updates of a single, randomly chosen perceptron carried out according to Eq.~\ref{eq:noisy_dynamics}.

Finally, for the system to work as a content-addressable memory, the $p$ patterns $\{\xi^\mu_i\}$ in Eq.~\ref{eq:hopfield_memories} should be rendered (meta)stable states of the network. To achieve this, the Hopfield model resorts to the Hebbian learning rule \cite{Hebb}, storing the patterns in the synaptic coefficients $J_{ij}$ between neurons by setting
\begin{equation}
    J_{ij} = \frac{1}{N} \sum_{\mu=1}^p \xi_i^\mu \xi_j^\mu.
    \label{eq:couplings_hopfield}
\end{equation}

\begin{figure}[b]
    \centering
    \includegraphics[width=\columnwidth]{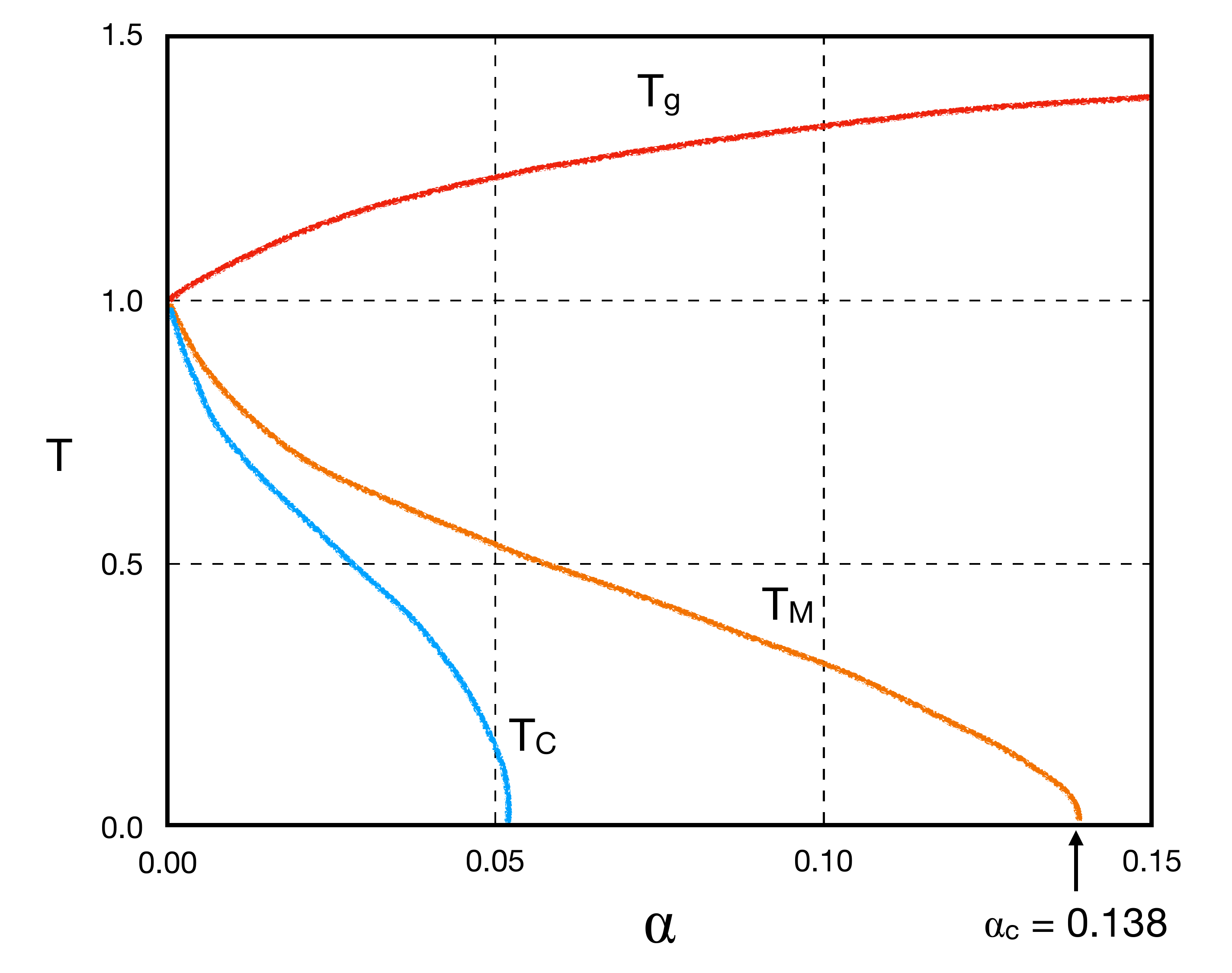}
    \caption{Sketch of the phase diagram of the Hopfield model in the thermodynamic limit $N\gg1$ as a function of the parameter $\alpha=p/N$ and the temperature $T$. Starting from the high-temperature disordered phase, $T_g$ marks the transition at which the system enters a spin glass phase; in both cases, the system is unable to recover the stored memory patterns. By lowering the temperature, memory retrieval states appear at $T_M$ as metastable states, and become minima of the free energy below $T_C$. Figure adapted from \cite{Amit_2_saturation}.}
    \label{fig:phase_diagram}
\end{figure}

Despite its relatively simple mathematical formulation, the Hopfield model exhibits an extremely rich phenomenology. Indeed, it has been proven \cite{Amit_2_saturation} that the efficacy of the system in retrieving the encoded memories along its dynamics is critically dependent on the amount of noise---that is, the temperature $T$ in Eq.~\ref{eq:noisy_dynamics}---and on the ratio $\alpha=p/N$ between the number of stored patterns and the size of the network. In the thermodynamic limit, the phase diagram of the system as a function of these parameters, sketched in Fig.~\ref{fig:phase_diagram}, is characterised by a paramagnetic and a spin glass phase (above the temperature curves labelled with $T_g$ and $T_M$, respectively) in which the network is not able to retrieve any memory, as well as a retrieval phase (below $T_M$) where the embedded patterns appear as thermodynamic metastable (below $T_M$ and above $T_C$) or stable (below $T_C$) states.

Given this general summary, we now describe the technical details associated with the numerical simulations of the Hopfield model carried out in this work. The three main simulation parameters are the network size $N$, the number of stored memory patterns $p$, and the system temperature $T$. Here, we focused on the study of individual realisations of the Hopfield model, characterising the properties of networks with given interaction matrices. The aim is that of establishing direct and system-specific relations between the behaviour of the mapping entropy as a function of the resolution, the features of the reduced representations, and the underlying model parameters.

We first investigated a small system consisting of $N=10$ neurons at $T = 0$, for which an exhaustive exploration of all the possible reduced representations of the system, see Sec.~\ref{subsec:mapping_entropy}, can be performed to extract the maximally informative ones. Subsequently, we addressed the analysis of larger networks with $N = 100$ at finite temperature. For fixed $N$ we performed simulations with $p$ ranging from $2$ to a maximum of $10$ memory patterns depending on the number of neurons in the ANN, See sec.~\ref{sec:results}. Furthermore, the temperature of the larger system was set to \textbf{$T=0.2$}, which, by assuming a finite-size phase diagram akin to the one presented in Fig.~\ref{fig:phase_diagram}, enables the network to fluctuate in its configurational space while remaining in the retrieval phase even at relatively high values of $\alpha$. Finally, for the system to explore all the memorised patterns more than once with nonzero probability, the pool of configurations employed in the analysis consisted of $n_{dyn}$ independent realisations of the dynamics of the network, in which each one started from a random assignment of the neuron states, and then eventually relaxed into an attractor basin. The number $n_{dyn}$ of independent trajectories was set to $1000$ for all the investigated values of $N$ and $p$, where the number of time cycles of each dynamics was chosen approximately equal to the retrieval period of the network. The time evolution of the system was obtained by relying on a stochastic asynchronous Glauber dynamics with a random updating sequence of the single perceptrons \cite{Amit_1, Amit_2_saturation, MBF, hopfield1982neural}.

\subsection{Mapping entropy optimisation workflow (MEOW)}
\label{subsec:mapping_entropy}

As discussed in the introduction, in this work we aim to identify simplified representations of a Hopfield network in terms of few ``important'' elemental units selected from a pool of variables that potentially contains also noisy, irrelevant ones. This is achieved only starting from a series of empiric observations of system states, without relying on previous knowledge of their underlying generative process. The simplification strategy that we employ in this endeavour is \emph{decimation}, that is, the analysis of the system in terms of a subset of its constituents and the implicit marginalisation on the discarded ones. Such a protocol has its origins in the process of coarse-graining (CG'ing) that lies at the heart of the renormalisation group in statistical and quantum physics, as well as of several modelling strategies of soft matter systems \cite{ref_8_HGP, ref_9_HGP, ref_10_HGP}. In the context of the Hopfield model, decimation consists of describing the network only in terms of a reduced number of selected neurons, a procedure that generally results in a loss of statistical information on the system. What we then look for, and identify with the aforementioned optimal simplified representations of the network, are the decimation mapping operators that, for a fixed number of retained neurons, provide a description of the ANN whose information content is \emph{as close as possible} to the original high-resolution reference. The protocol implementing this strategy constitutes the mapping entropy optimization workflow (MEOW) recently developed by some of us \cite{JCTC, HGP, giulini2024excogito}; we now briefly summarise the strategy underlying the MEOW and the associated fundamental information-theoretical ingredients, focussing our attention on the technical details related to the application of this workflow to the resolution reduction of a Hopfield network.

Consider a dynamic trajectory of a Hopfield ANN as obtained via the simulation protocol described in Sec.~\ref{subsec:Hopfield_model}: at each time $t$, the high-resolution or ``fine-grained'' configuration of the system is given by the state of all of its constituent neurons, $\phi(t)=(\sigma_1(t),\sigma_2(t),...,\sigma_N(t))$ with $\sigma_i(t)\in\{-1,1\}$. From such trajectory (or an ensemble of them), one can construct the \emph{empirical} probability $p(\phi) = p({\sigma_1,\sigma_2,...,\sigma_N})$ of observing a specific microstate $\phi$, which is nothing but the frequency with which the selected configuration appears in the time series, namely
\begin{equation}
    p(\phi)=\frac{1}{T}\sum_{t=1}^T\prod_{i=1}^N\delta(\sigma_i,\sigma_i(t)),
    \label{eq:empirical_prob}
\end{equation}
where $T$ is the total number of simulation steps.\footnote{We stress that the empirical nature of $p(\phi)$ is such that, although in principle the possible microstates of the network are $2^N$, some of these configurations will not be visited along the trajectory, either as a mere consequence of the finiteness of the sample or due to the additional presence of ergodicity-breaking phenomena, see Fig.~\ref{fig:phase_diagram}. In the following, such missing configurations will be \emph{excluded} from the high-resolution state space rather than being endowed with a vanishing empirical probability. This procedure is akin to the one of taking restricted equilibrium averages in systems displaying ergodicity breaking; furthermore, it is particularly suited in the analysis of time series for which the properties of the configurational space and/or the mechanism underlying the generation of the high-resolution samples are not known \cite{Multiscale_relevance_Marsili}.}

By explicitly accounting for the state of all constituent neurons, the empirical fine-grained probability $p(\phi)$ provides a complete characterisation of the (observed) statistical properties of the network. The complexity inherent to such a high-dimensional description, however, can hinder the process of distilling the relevant information on the system out of a potentially noisy background; it is thus natural to investigate whether simplified representations of the network can be constructed that are capable of enhancing the signal-to-noise ratio. As previously introduced, the elemental ingredient lying at the core of MEOW to tackle this problem is a coarse-graining procedure that decimates the $N$ high-resolution degrees of freedom of the network, describing the latter only in terms of a subset of $n_{cg}<N$ neurons. Let us first discuss how such an operation is implemented in practice for a \emph{specific} selection of the sites, and what are its implications on our knowledge of the global statistical properties of the system. Starting from this, we will then describe the identification of the aforementioned maximally informative reduced representations of the network.

We perform an analysis of the system in which only a specific subset of neurons out of the $N$ constituent ones is explicitly accounted for. Practically, this amounts to introducing a mapping operator $M(\phi)$ that projects each high-resolution configuration $\phi=({\sigma_1,\sigma_2,...,\sigma_N})$ of the ANN onto its low-resolution counterpart $\Psi_{\phi} = M(\phi) \equiv (\sigma_{i_1},\sigma_{i_2},...,\sigma_{i_{n_{cg}}})$; the latter consists of only the states of the $n_{cg}$ neurons ($i_1,..., i_{n_{cg}}), \ i_{\nu}\in \{1,.., N\}$, that were selected to be retained in the low-resolution representation. Critically, this procedure reverberates on the statistical properties of the network \emph{as inspected in its decimated form}: the mapping operator induces a \emph{low resolution} empirical probability of observing a specific decimated configuration $\Psi$, $P(\Psi)$, that can be obtained from $p(\phi)$ as
\begin{equation}
    \label{eq:cg_prob}
    P(\Psi)=\sum_{\phi}p(\phi)\delta(\Psi,\Psi_{\phi}),
\end{equation}
namely as the marginalised high-resolution probability of all the fine-grained states that map on the selected low-resolution configuration. From $P(\Psi)$ all the properties of the reduced network can be obtained; at the same time, the effect of the projection is to conceal the detailed features that pertain to the set of integrated neurons, and only a partial description of the system remains available. One could then ask to what extent an observer provided with such limited knowledge could succeed in deducing from it the same features encoded in the high-resolution reference.

The goal is hence to reconstruct the statistical properties of the full network, namely the fine-grained empirical probability $p(\phi)$, \emph{only given} its reduced counterpart $P(\Psi)$. In doing so, we note that, after mapping, no additional information on each microstate is readily accessible but for its association with the corresponding low-resolution label; a reversal of the decimation procedure should thus be compatible with a maximum entropy principle in which all the microstates that map onto the same low-resolution configuration are attributed equal likelihood to occur. The resulting reconstructed or \emph{backmapped} high-resolution probability distribution, $\bar{p}(\phi)$, accordingly reads
\begin{equation}
    \label{eq:backmapped_prob}
    \bar{p}(\phi)=\frac{P(\Psi_{\phi})}{\Omega(\Psi_{\phi})},
\end{equation}
where 
\begin{equation}
    \label{eq:cg_degeneracy}
    \Omega(\Psi) = \sum_{\phi'} \delta(\Psi,\Psi_{\phi'})
\end{equation}
is the degeneracy of decimated configuration $\Psi$, that is, the observed number of unique fine-grained states of the network that map onto $\Psi$. From Eq.~\ref{eq:backmapped_prob} it follows that describing the network in terms of a selected subset of its neurons has generated a \emph{loss of statistical information on the system}, in that, in contrast to the high-resolution reference $p(\phi)$, upon backmapping all the microstates of the ANN that compose each low-resolution configuration $\Psi$ have become statistically equivalent; additionally, we observe that the reconstructed probability $\bar{p}(\phi)$, which is common to all high-resolution microstates mapping onto $\Psi$, is given by the average of their original probabilities.

The loss of information generated by coarsening the representation of the network can be quantified, in information-theoretic terms, \emph{via} the \emph{mapping entropy} $S_{map}$ \cite{JCTC,HGP,Shell_Scott,rudzinski2011coarse,noid2023perspective, giulini2024excogito}, with
\begin{eqnarray}\label{eq:mapping_entropy}
    &&S_{map} = \sum_\phi p(\phi)\ln \left[\frac{p(\phi)}{\bar{p}(\phi)} \right]= \nonumber \\ 
    &&= \sum_\phi p(\phi)\ln \left[ p(\phi) \frac{\Omega(\Psi_\phi)}{P(\Psi_\phi)} \right].
\end{eqnarray}
Being a Kullback-Leibler divergence between the original and the backmapped probability distributions \cite{kullback1951information}, the mapping entropy is nonnegative because of Gibbs' inequality, with $S_{map}=0 \Longleftrightarrow \bar{p}(\phi)=p(\phi) \ \forall \phi$. Most importantly, since the detailed form of $\bar{p}(\phi)$---and consequently the resulting mapping entropy---only depends on the specific subset of neurons chosen to represent the network at low resolution, see Eqs.~\ref{eq:cg_prob},~\ref{eq:backmapped_prob} and~\ref{eq:cg_degeneracy}, it follows that different mapping operators can be associated with different amounts of information loss on the statistical properties of the reference system. One is then naturally led to look, in the space of possible selections of neurons, for those that \emph{minimise} $S_{map}$; a low value of mapping entropy implies that, in spite of the resolution loss of the system representation, the information content available from the retained neurons is close to the one encoded in the original distribution. These mappings constitute the so-called \emph{maximally informative} reduced representations that can be designed to investigate the behaviour of the system, and their identification and analysis is at the core of the strategy implemented in MEOW \cite{JCTC,errica2021deep,HGP,giulini2024excogito}.

To examine the properties of a Hopfield network through the lenses of the mapping entropy, we rely on a code recently developed by some of us, the extensible coarse-graining toolbox, or EXCOGITO \cite{giulini2024excogito}. Specifically, the program takes as input the fine-grained probabilities of all the configurations explored in the course of a series of simulations of the Hopfield model, where the probability of a given microstate $p(\phi)$ is calculated as the empirical frequency of its occurrence in the dataset, see Eq.~\ref{eq:empirical_prob}. The software then proceeds to identify the maximally informative reduced representations of the network by determining, among the possible selections of a fixed number $n_{cg}$ of neurons, the ones that minimise $S_{map}$; this procedure is then iterated for various levels of resolution $n_{cg}$. For each analysed mapping, its value of $S_{map}$ is calculated by clustering all the fine-grained configurations in which the selected neurons are in the same state, enabling the reconstruction of the empirical probability distribution $P(\Psi)$ of the decimated representation, as well as of the degeneracy factor $\Omega(\Psi)$ (Eqs.~\ref{eq:cg_prob}-\ref{eq:cg_degeneracy}) that enters in the definition of the mapping entropy (Eq.~\ref{eq:mapping_entropy}). Critically, we note that the overall network size $N$ plays a crucial role in fulfilling the optimisation of $S_{map}$. Indeed, for small values of $N$ all the possible $n_{map}=2^N-1$ mappings of the system can be exhaustively probed (and ranked) to detect the maximally informative ones at each degree of coarse-graining $n_{cg}$; on the contrary, such an extensive exploration becomes rapidly unfeasible as the number of constituent neurons increases \cite{ref_25_HGP}. In this latter case, the software minimises $S_{map}$ in the space of the possible reduced representations of the network that can be constructed at a given $n_{cg}$ \textit{via} a Monte Carlo simulated annealing (SA) procedure \cite{ref_62_JCTC, ref_63_JCTC}; the algorithm proposes a transition from one neuron subset to another other, the two differing by one single retained unit, and the new subset is accepted or rejected according to a Metropolis-like criterion that employs $S_{map}$ as cost function. By exponentially decreasing the SA effective temperature parameter along the course of the simulation, the latter is gradually pushed to identify (local) minima of the mapping entropy. For each $n_{cg}$ a series of $K_{\text{SA}}=48$ independent SA simulations were performed, thus resulting in the detection of \emph{a pool} of maximally informative reduced representations of the $N=100$ network at each degree of coarse-graining.

Finally, in the following the mapping entropy is studied as a function of the \emph{resolution} of the simplified representations \cite{Marsili_1}, which is defined as:
\begin{equation}
    H_S = - \sum_\Psi P(\Psi) \ln P(\Psi).
    \label{eq:resolution}
\end{equation}
This quantity, which technically is the Shannon entropy of the empirical low-resolution probability distribution, was introduced by Marsili and coworkers as a measure of the degree of detail with which an empirical dataset is described; a qualitative understanding of this quantity can in fact be gathered by observing that, if all elements of a dataset of size $\mathcal M$ are labelled differently, their empirical probability is $1/\mathcal M$, which returns the largest value of the resolution ($\ln \mathcal M$); by grouping elements e.g. through some clustering procedure, the empirical probability of each group, defined as the fraction of elements in it, is associated to a lower value of the entropy. The lowest value is attained when all elements are included in the same cluster, which corresponds to a null entropy. We redirect the reader interested in this topic to the available literature \cite{Criticality_Informative_samples_Marsili,Marsili_1,Multiscale_relevance_Marsili,On_sampling_and_modeling_Marsili,Mele_Potestio,HGP}.

\begin{figure*}
    \subfloat[$p=1$]{\includegraphics[height=7.0 cm]{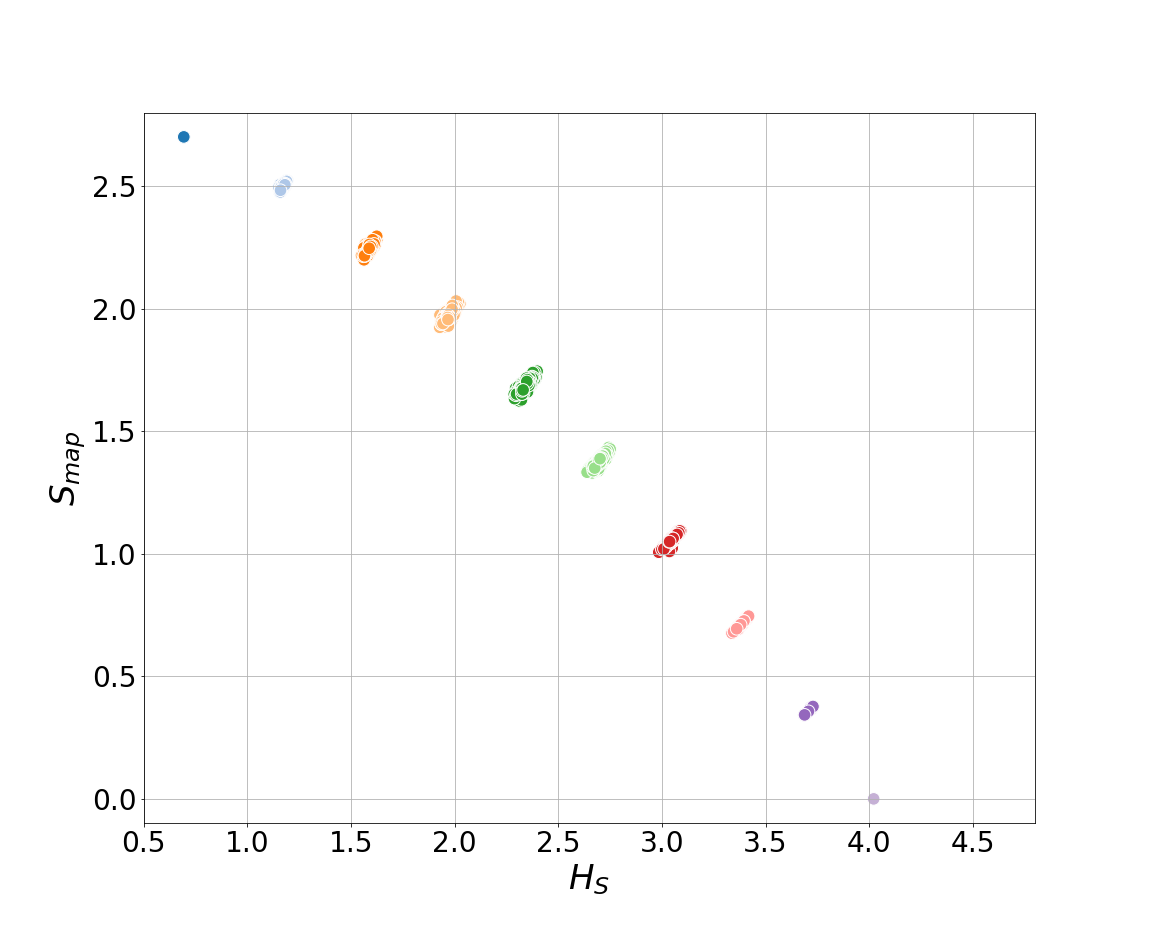}}\hfill
    \subfloat[$p=2$]{\includegraphics[height=7.0 cm]{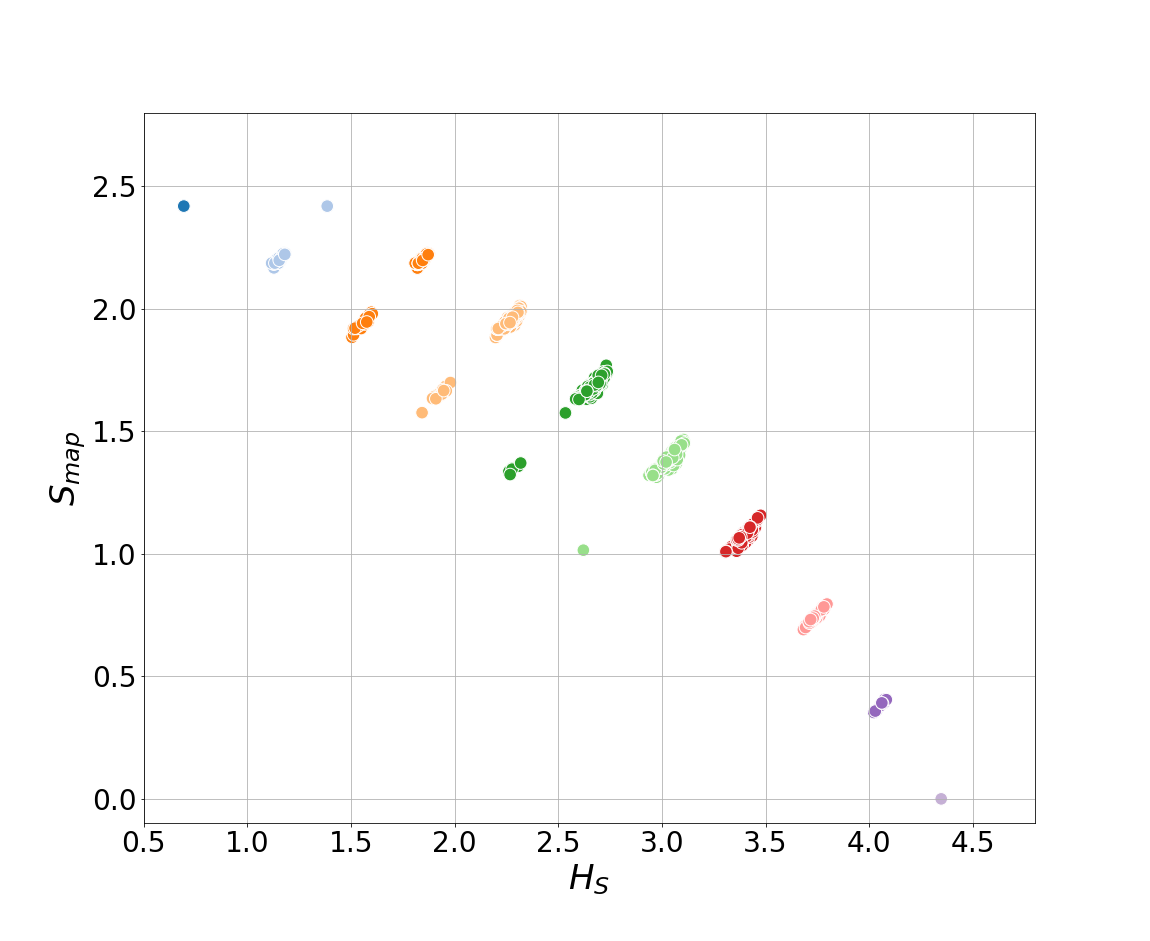} \label{N10_p2}}\\[-3ex]  %%<-- in this line
    \subfloat[$p=3$]{\includegraphics[height=7.0 cm]{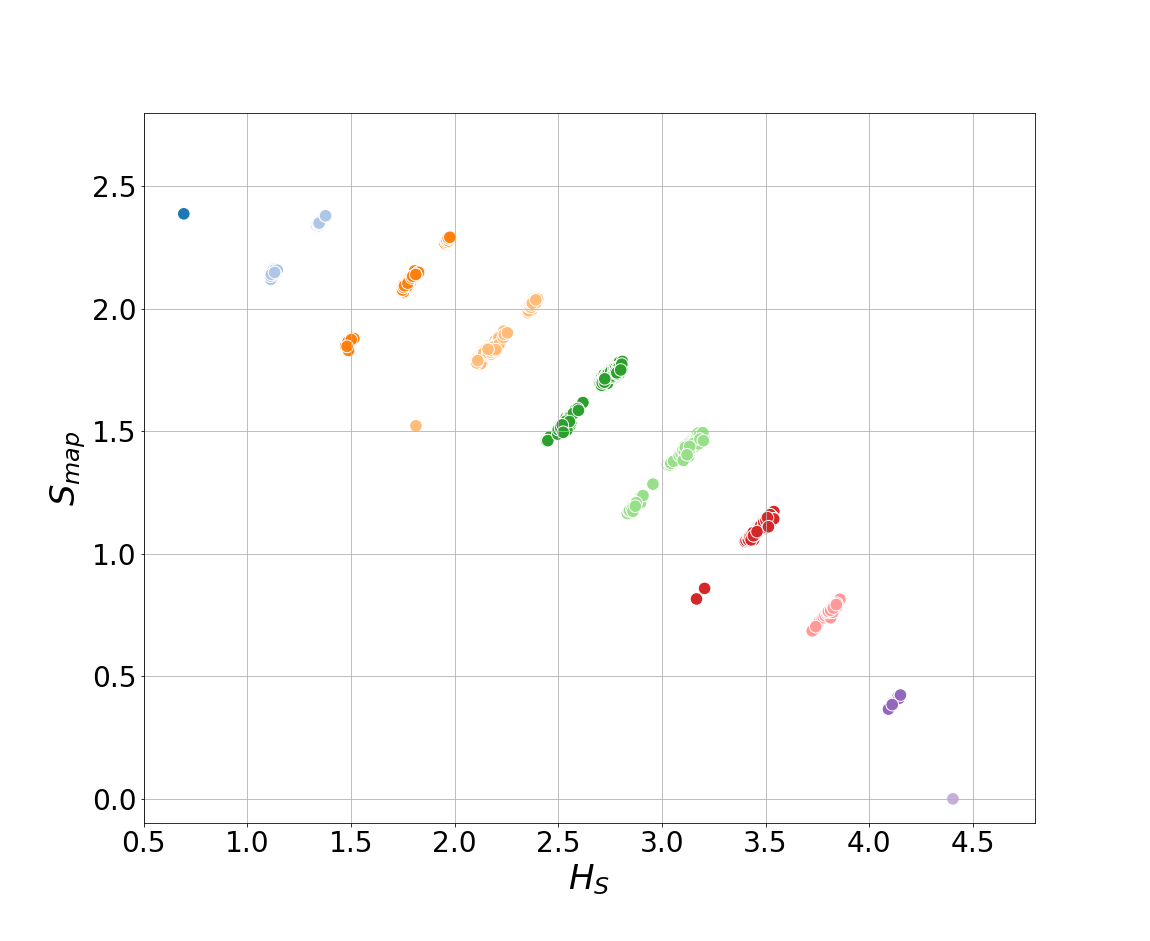} \label{fig:Smap_total_N10_p3}}\hfill
    \subfloat[$p=4$]{\includegraphics[height=7.0 cm]{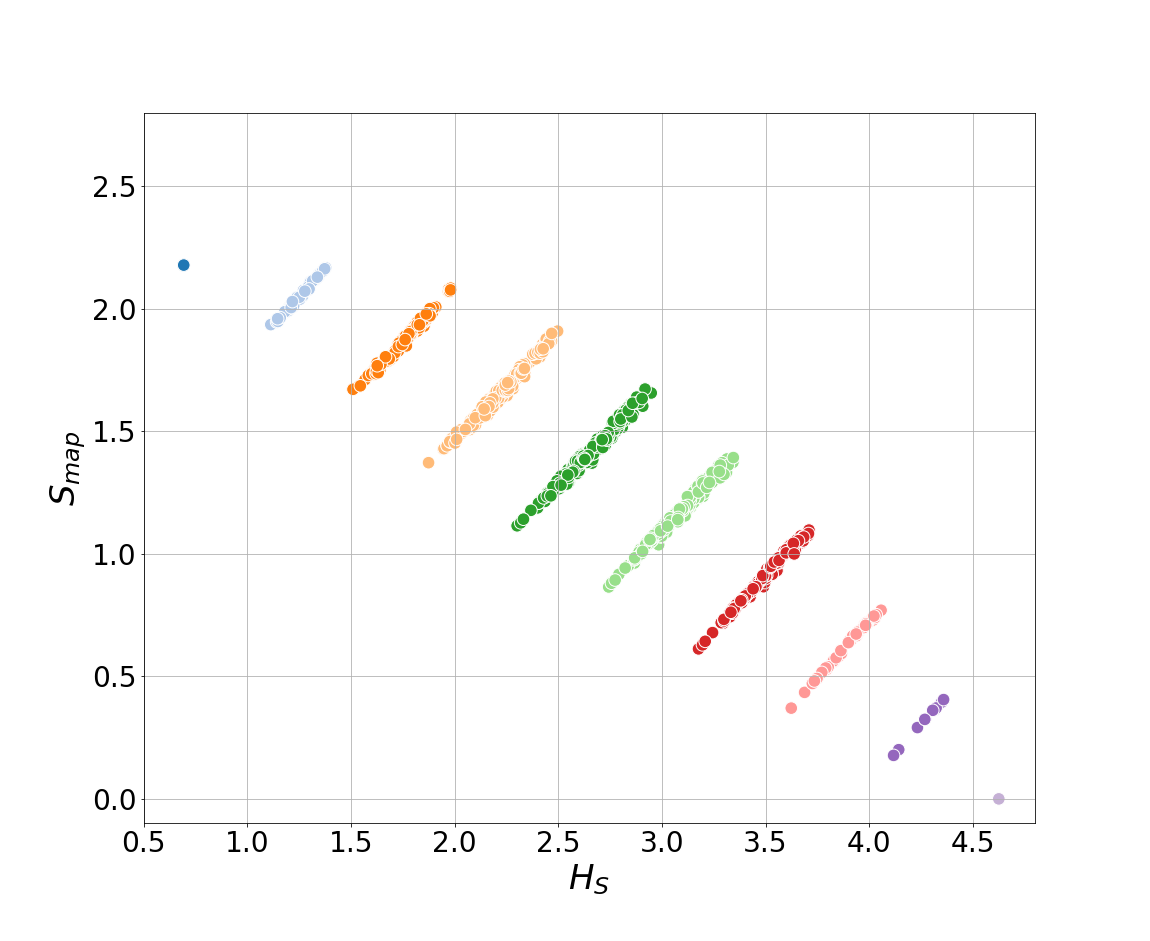}}\\[-3ex]
    \begin{minipage}[c]{0.5\textwidth}
    \subfloat[$p=5$]{\includegraphics[height=7.0 cm]{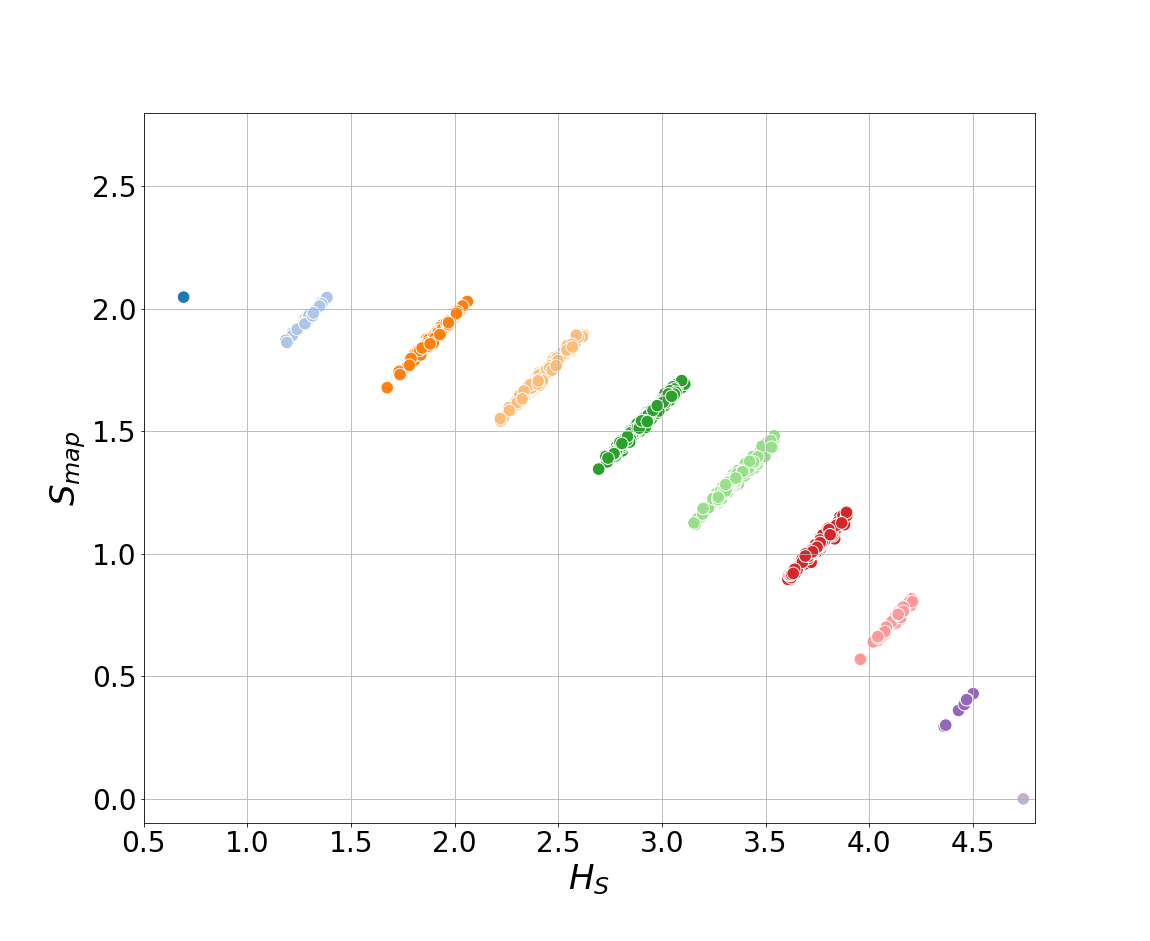}}
    \end{minipage}\hfill
        \begin{minipage}[c]{0.45\textwidth}
            \begin{center}
            \subfloat{\includegraphics[width=4.5 cm]{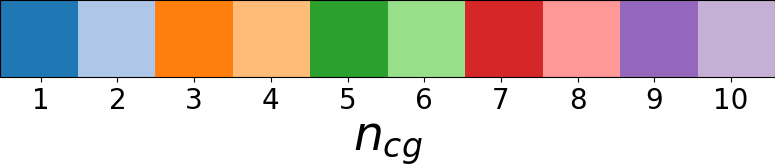}}
            \end{center}
        \caption{Mapping entropy of the $N=10$ Hopfield model plotted as a function of resolution for five values of $p$. All possible simplified representations of the system are covered for numbers of retained neurons $n_{cg}$ ranging from $1$ to $10$. Simulations have been carried out with parameters $n_{dyn}=1000$, $\tau = 3$, $T=0$ (see main text for further details).}
        \label{fig: Smap_N10_total}
    \end{minipage}
\end{figure*}

\section{Results and Discussion}\label{sec:results}

\subsection{Matrix reconstruction and bias detection for a N=10 Hopfield model}

In this section we report the results of the analysis of a relatively small network, with the aim of building an intuition of the properties of the mapping entropy of optimal low-resolution representations in the context of a fully-controllable system. Specifically, the key feature of this model is the viability of exhaustively enumerating all possible decimated mappings one can employ to study the network. Hence, we can identify the absolute minima of $S_{map}$ as a function of the number of retained neurons and inspect the corresponding mappings. Furthermore, we illustrate a strategy to approximately reconstruct the interaction matrix underlying the model from an analysis of the optimal decimated representations.

We thus begin to examine the properties of the Hopfield model through the lenses of mapping entropy considering a specific realisation of a small network of $N= 10$ at $T = 0$, with an amount of randomly-generated stored patterns in the range $p\in\{1,..,5\}$. As anticipated, the relatively small total number of reduced representations (irrespectively of the resolution level $n_{cg}$) that can be designed to describe a network of this size, $n_{map} = 2^{10} - 1=1023$, allows us to perform an exhaustive exploration of all of them for every value of $p$ investigated; the state configurations sampled by each setup in the course of its simulation were provided as input to the EXCOGITO software for the MEOW analysis. The obtained results will thus constitute a benchmark for the following analysis on a larger system discussed in Sec.~\ref{sec:n100network}.

\subsubsection{Properties of the minima of the mapping entropy and their relation with the memory patterns}

The mapping entropy and resolution associated with each reduced representation of the system were calculated as detailed in Sec.~\ref{subsec:mapping_entropy}, providing, at fixed number of stored patterns, a point in the corresponding $S_{map} \ vs. \ H_{S}$ plane; combined together, these results fully characterise the landscape of information loss attained while observing a simple, yet nontrivial Hopfield network in terms of a subset of its constituent units.

The $S_{map}\ vs. \ H_S$ graphs arising from this analysis are presented in Fig.~\ref{fig: Smap_N10_total}, where we report results separately for each value of $p$ and further label decimation mappings according to their degree of detail $n_{cg}$. Overall, two features can be readily appreciated: first, and as expected, the mapping entropy on average decreases as the number of retained neurons increases, vanishing for $n_{cg}=N$; this is due to the fact that the larger the number of elements considered in the reduced description, the smaller the amount of information that is lost on the statistical properties of the high-resolution reference.

Second, it appears that a linear relation holds between $S_{map}$ and $H_S$ at a fixed $n_{cg}$. This behaviour can be rationalised through the definition of the mapping entropy provided in Eq.~\ref{eq:mapping_entropy}, which can be rewritten as
\begin{equation}
    \begin{split}
        S_{map} = -H_{S}^\phi + H_{S} + \sum_\Psi P(\Psi) \ln (\Omega(\Psi)).
    \end{split}
    \label{Eq:Smap-Resolution}
\end{equation}

In Eq.~\ref{Eq:Smap-Resolution}, $H_{S}^\phi = -\sum_{\phi}p(\phi)\ln p(\phi)$ is the fine-grained resolution of the reference model, a constant quantity that is independent of the mapping. On the contrary, $H_S$ is strictly related to the choice of the reduced representation, hence contributing to variations in $S_{map}$. We note that this in principle also holds for the third factor in Eq.~\ref{Eq:Smap-Resolution}, in which the logarithm of the degeneracy $\Omega(\Psi)$ of the decimated representation labels is averaged over the corresponding probability distribution; for a sufficiently large sample of high-resolution configurations, however, this term only depends on the mapping \textit{via} the amount of retained sites $n_{cg}$, and is thus constant at a fixed degree of detail \cite{HGP}. In such limit, Eq.~\ref{Eq:Smap-Resolution} thus clarifies the linear dependence of $S_{map}$ on $H_{S}$ for fixed value of $n_{cg}$ observed in Fig.~\ref{fig: Smap_N10_total}, according to which the lower the resolution of the reduced representation, the lower its information loss.

\begin{figure}[]
    \subfloat[$ $]{\includegraphics[width=7 cm]{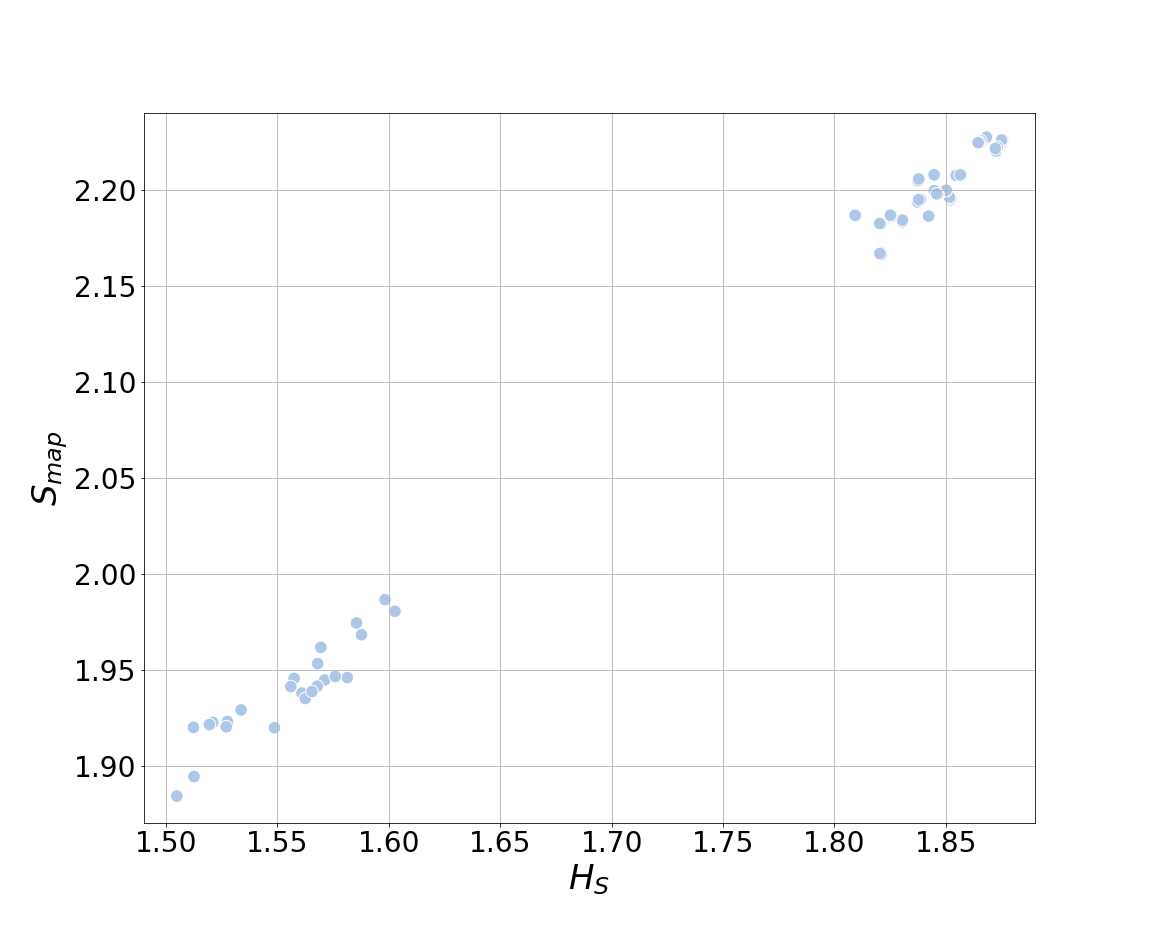} \label{fig:close_up_Smap_N10_a}}\hfill
    \subfloat[$ $]{\includegraphics[width=7 cm]{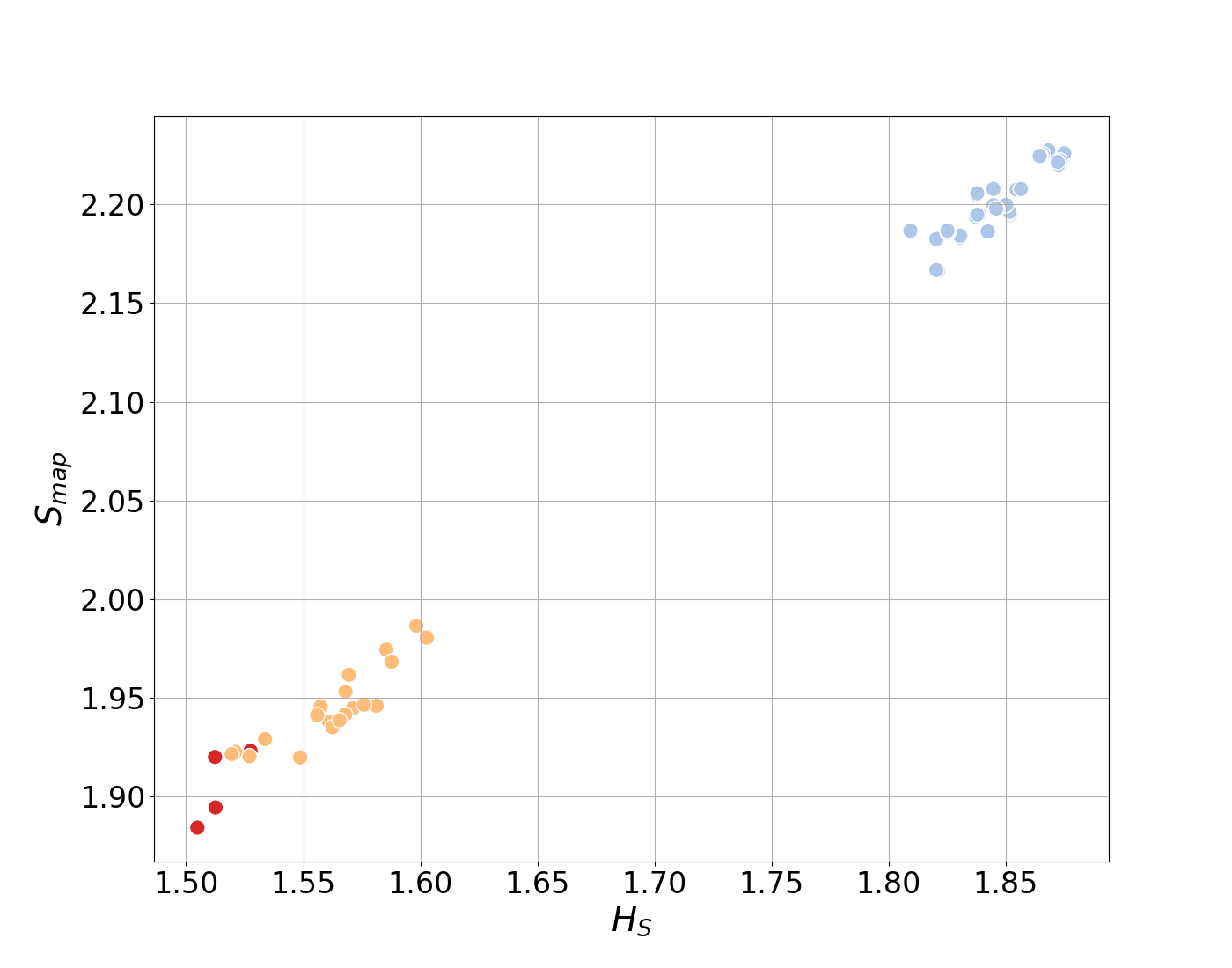} \label{fig:close_up_Smap_N10_b}}\hfill
    \subfloat[$ $]{\includegraphics[width=7 cm]{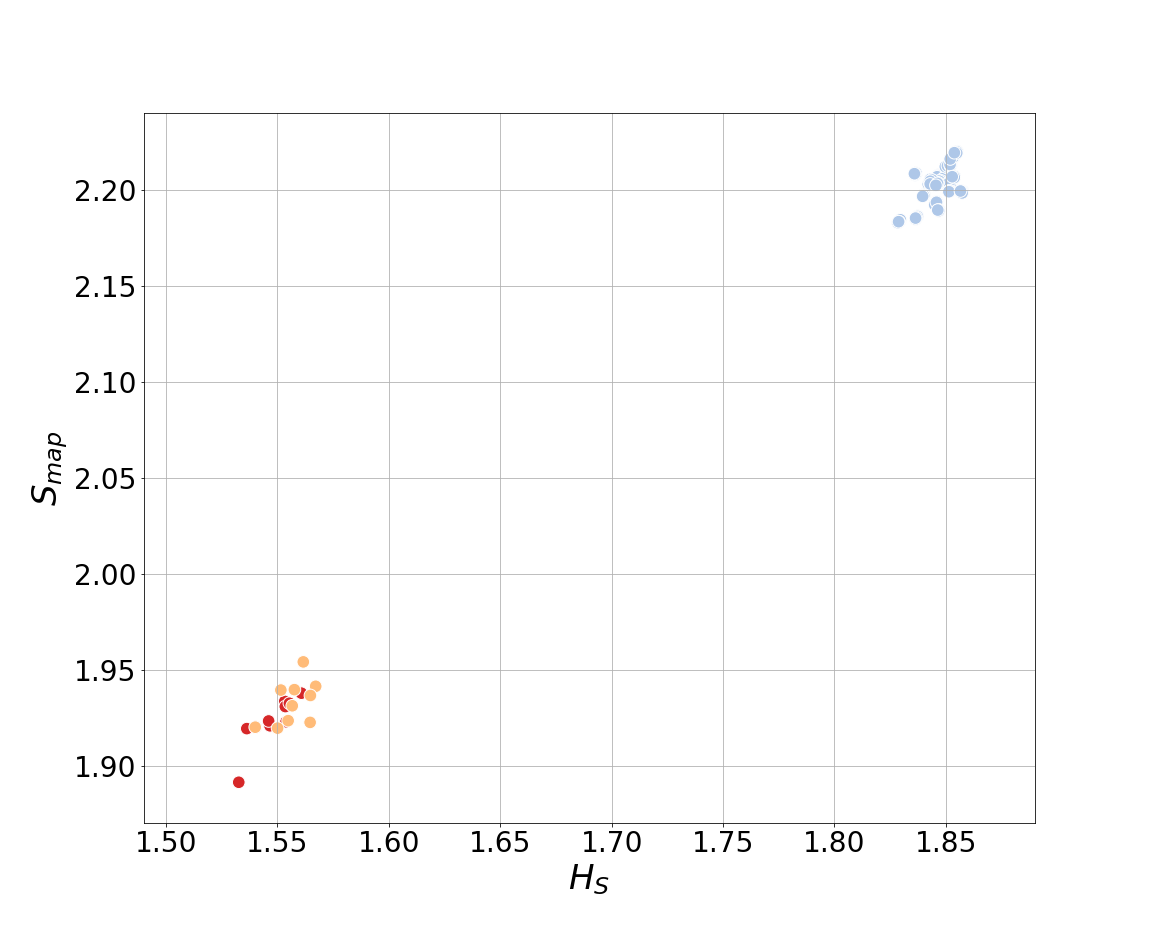} \label{fig:close_up_Smap_N10_c}}
    \caption{Panel (a): close-up of Fig. \ref{N10_p2}, showing the mapping entropy for mappings with $n_{cg} = 3$. Panel (b): same as panel (a), where mappings are coloured depending on the properties of the retained neurons. Red dots: mappings that contain only neurons whose memories have the same value on all patterns. Orange dots: mappings that contain only neurons whose memories are opposite. Light blue dots: mappings that contain neurons coming from both latter groups. Panel (c): mapping entropy for mappings with $n_{cg} = 3$ of an $N=10$ Hopfield model with biased patterns, see Eq. \ref{eq:biased_patterns_p2}. Red dots: mappings that retain only spins from the first five (identical among the patterns). Orange dots: mappings that retain only spins from the last five. Light blue dots: mappings that retain spins from both the first and the second half.}
    \label{fig:close_up_Smap_N10}
\end{figure}

At the same time, by further analysing the behaviour of the mapping entropy as a function of the resolution at a fixed degree of detail, a closer inspection of  Fig.~\ref{fig: Smap_N10_total} reveals that, despite the overall linearity of $S_{map}$ in $H_S$, for several values of $n_{cg}$ the reduced representations of the system split into distinct clusters separated by gaps in mapping entropy. This feature is particularly evident in the case of $p=2$ and $p=3$ (resp. Fig.~\ref{N10_p2} and~\ref{fig:Smap_total_N10_p3}), while it is almost absent for the other amounts of stored patterns. One is then naturally led to ask why such a  ``classification'' of decimation mappings appears; critically, the reason for this is to be found in the structure of the specific realisation of the synaptic weight matrix $J_{ij}$ characterising the interactions among the network's constituents. To better clarify this point, we focus on the simplest case with $p=2$ retaining $n_{cg}=3$ sites---for which a zoomed version of the results presented in Fig.~\ref{N10_p2} is reported in Fig.~\ref{fig:close_up_Smap_N10}; furthermore, we first consider a Hopfield network with two ``biased'' patterns, namely
\begin{equation}
    \begin{tabular}{ ccccccccccc }
     $\{\xi^1\}$ = & \textbf{-1} & \textbf{-1} & \textbf{-1} & \textbf{-1} & \textbf{-1} & -1 & -1 & 1 & -1 & 1  \\ 
     $\{\xi^2\}$ = & \textbf{-1} & \textbf{-1} & \textbf{-1} & \textbf{-1} & \textbf{-1} & 1 & 1 & -1 & 1 & -1
    \end{tabular}
    \label{eq:biased_patterns_p2}
\end{equation}

In Eq.~\ref{eq:biased_patterns_p2} the first five neurons have equal memory values in both patterns, while the memories of the last five are opposite in sign. As a consequence of the Hebbian rule in Eq.~\ref{eq:couplings_hopfield}, if the product $\xi^{\mu}_i\xi^{\mu}_j$ of the memories of a pair of spins $(\sigma_i,\sigma_j)$ is equal in all the stored patterns, the modulus of their coupling achieves its highest possible value ($|J_{ij}|=p/N$ in the general case of $p$ patterns), thus rendering these neurons maximally interacting. If this does not hold, cancellations occur that can also result in a complete decoupling of the pair. In particular, the memory patterns in Eq.~\ref{eq:biased_patterns_p2} generate the block-diagonal synaptic weight matrix reported in Eq.~\ref{S_biased}, in which neurons end up being partitioned into two groups: units in each group maximally interact with each other (with $|J_{ij}|=2/10$ for $p=2$), while units from different groups do not. The first block consists of the first five spins in Eq.~\ref{eq:biased_patterns_p2} that have equal memories in both patterns, while the other block pertains to the remaining ones.
\begin{equation}
    S=  
    \begin{pmatrix}
    0& 0.2&0.2&0.2&0.2&0&0&0&0&0\\
    0.2&0&0.2&0.2&0.2&0&0&0&0&0\\
    0.2&0.2&0&0.2&0.2&0&0&0&0&0\\
    0.2&0.2&0.2&0&0.2&0&0&0&0&0\\
    0.2&0.2&0.2&0.2&0&0&0&0&0&0\\
    0&0&0&0&0&0&0.2&$-0.2$&0.2&$-0.2$\\
    0&0&0&0&0&0.2&0&$-0.2$&0.2&$-0.2$\\
    0&0&0&0&0&$-0.2$&$-0.2$&0&$-0.2$&0.2\\
    0&0&0&0&0&0.2&0.2&$-0.2$&0&$-0.2$\\
    0&0&0&0&0&$-0.2$&$-0.2$&0.2&$-0.2$&0
    \label{S_biased}
    \end{pmatrix}
    .
\end{equation}

The mapping entropy analysis performed on all the possible reduced representations that can be designed for this biased Hopfield network by retaining $n_{cg}=3$ neurons is presented Fig.~\ref{fig:close_up_Smap_N10_c}. Also in this case we observe the appearance of two clusters separated by a gap in $S_{map}$; by investigating their composition, it emerges that the low-resolution representations that enter the group of lower $S_{map}$ are the ones that only contain spins coming both from either the first or the second blocks in Eq.~\ref{S_biased}, thus consisting of retained neurons that are maximally interacting among themselves and decoupled with the reminder. Constructing ``mixed'' decimation mappings that gather interacting as well as noninteracting spins results in a higher information loss on the properties of the original network. This result, obtained with a model crafted \emph{ad hoc}, is consistently found in the $p=2$, random patterns model of Fig.~\ref{N10_p2}, see Fig.~\ref{fig:close_up_Smap_N10_b}. In fact, also in this case the mappings belonging to the lower $S_{map}$ cluster are those retaining a subset of spins whose memory states are either identical or are opposite in sign between the two patterns.

\begin{figure*}[t]
    \subfloat[$p=1$]{\includegraphics[height=4.5 cm]{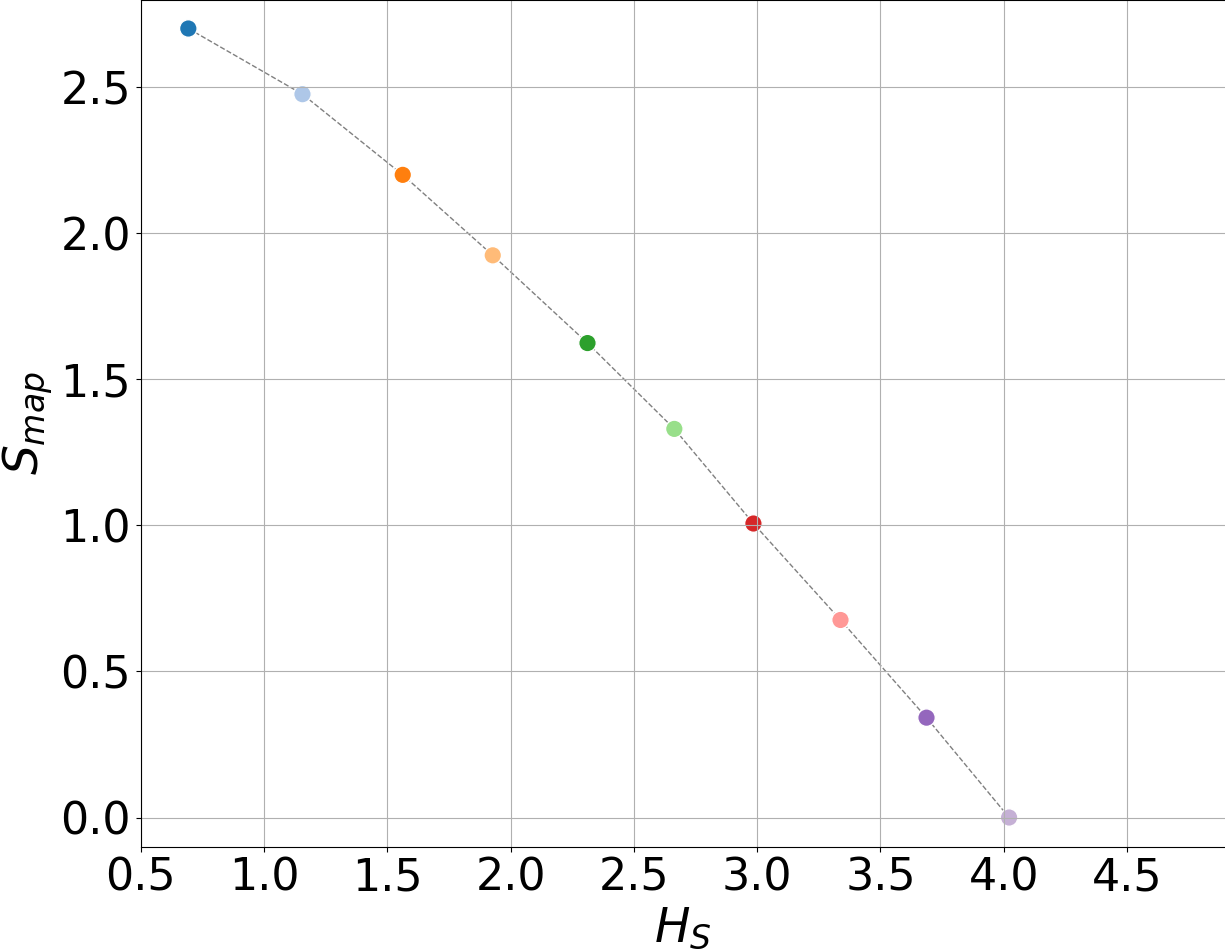}}\hfill
    \subfloat[$p=2$]{\includegraphics[height=4.5 cm]{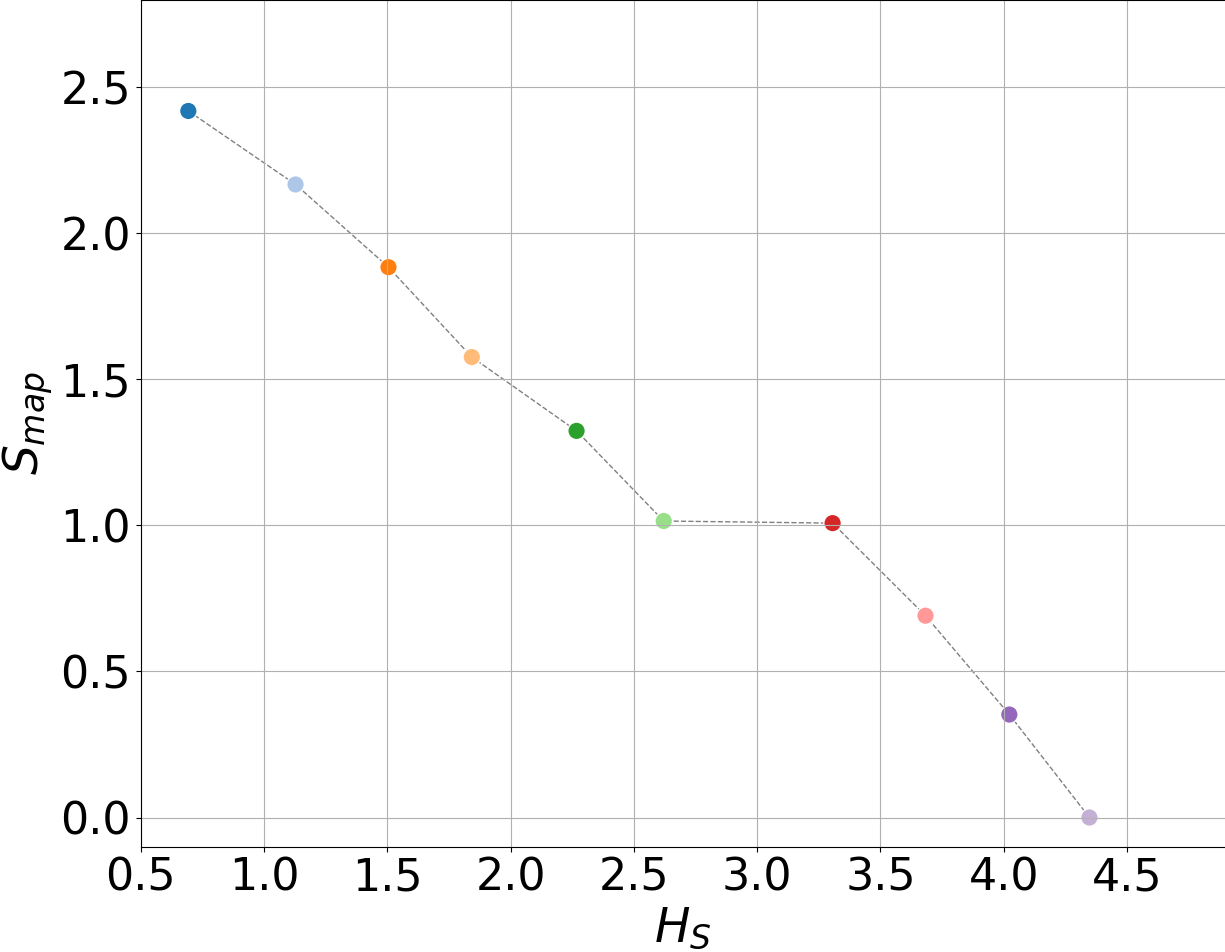}}\hfill
    \subfloat[$p=3$]{\includegraphics[height=4.5 cm]{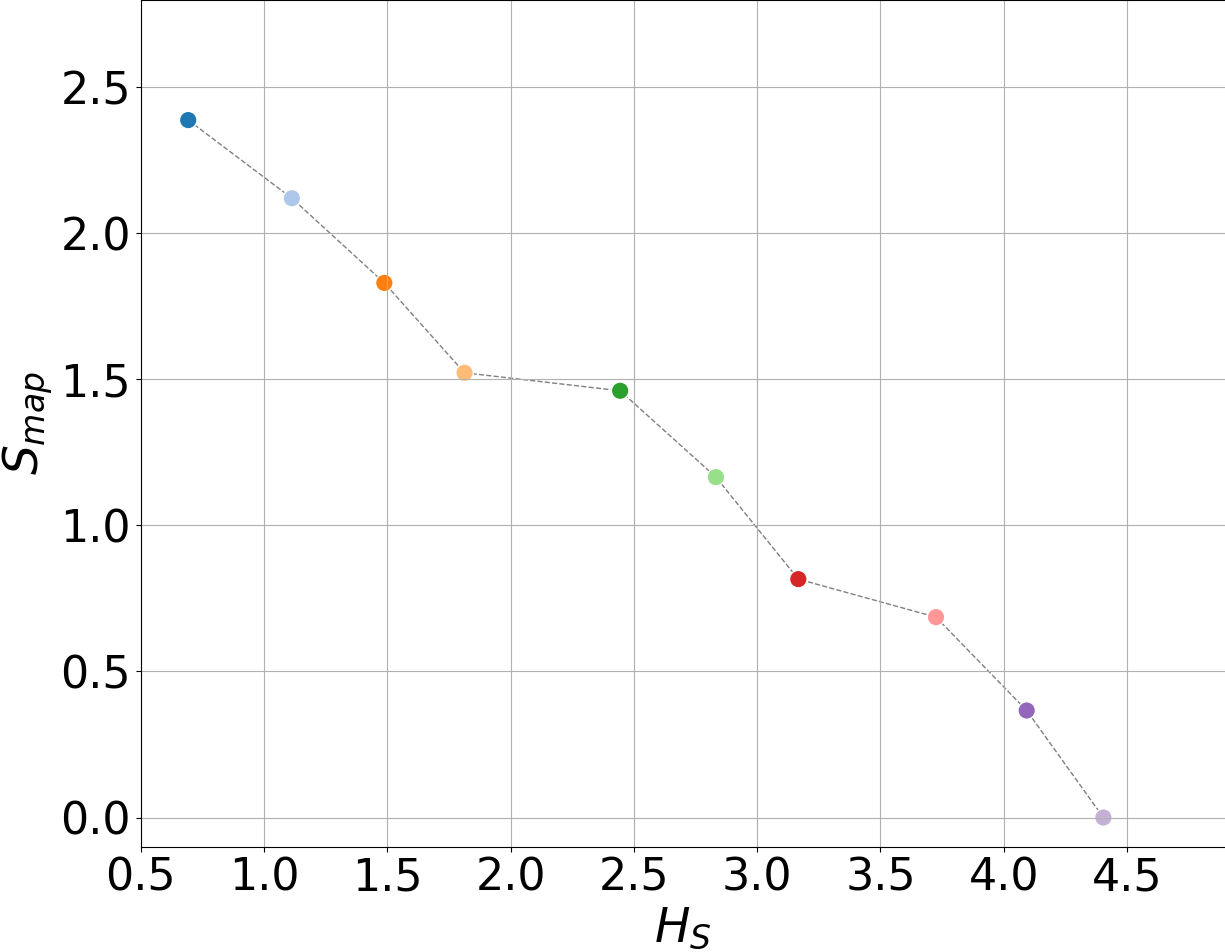}}\\
        \begin{minipage}[c]{0.67\textwidth}
            \subfloat[$p=4$]{\includegraphics[height=4.5 cm]{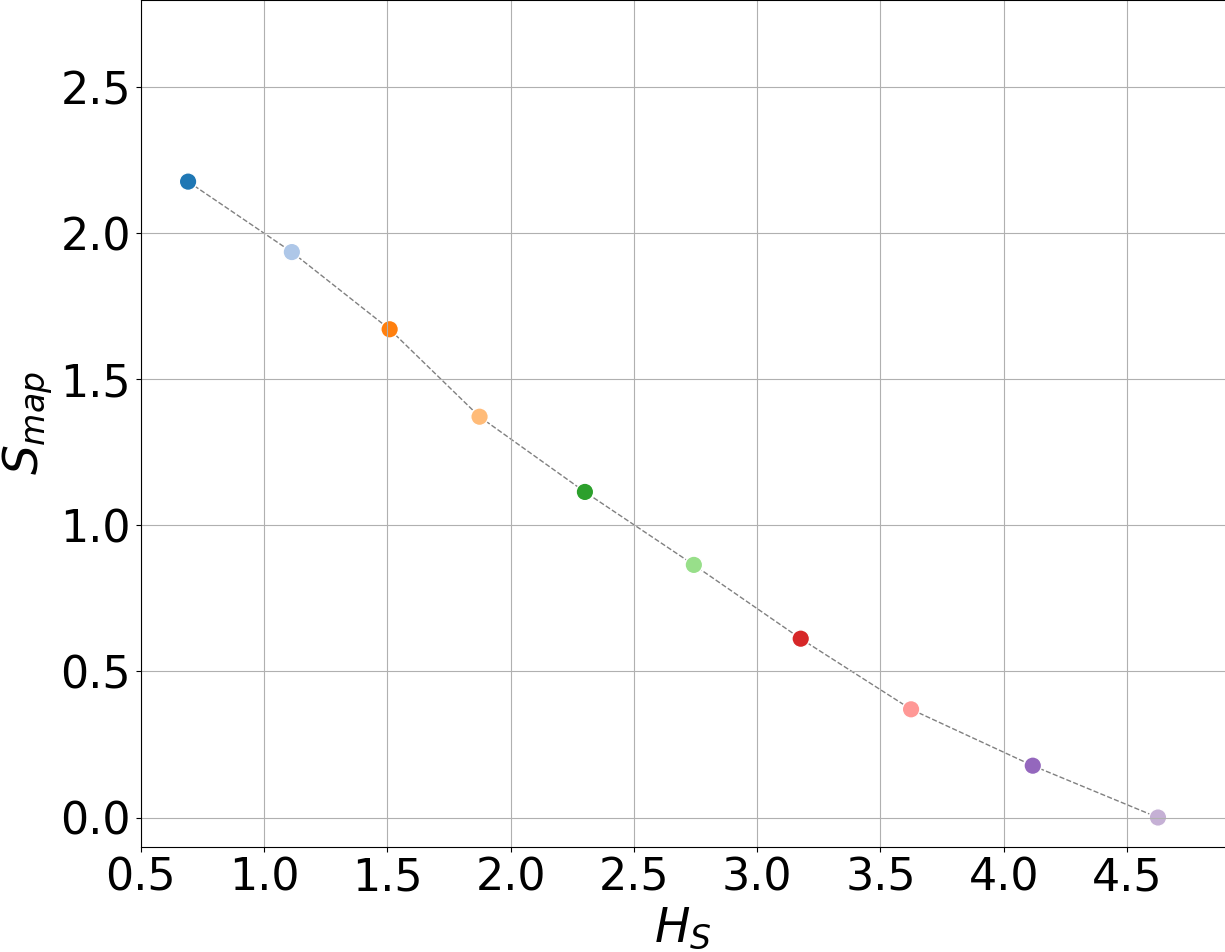}}\hfill
            \subfloat[$p=5$]{\includegraphics[height=4.5 cm]{figures/N10_Smap/2_Min_N10_p1.png}\label{fig:MIN_N10_p5}}
        \end{minipage}\hfill
        \begin{minipage}[c]{0.32\textwidth}
            \caption{Mapping entropy plotted as a function of resolution for $p = 1, \cdots \ 5$, for the coarse-grained representations of the system that minimize the loss of information at fixed $n_{cg}$. The graphs refer to the same model studied in Fig.~\ref{fig: Smap_N10_total}. }
            \label{fig: Smap_N10_MIN}
            \begin{center}
                \subfloat{\includegraphics[width=4.5 cm]{figures/Legend/N10.png}}
            \end{center}
        \end{minipage}
\end{figure*}

\begin{figure*}
    \subfloat{\includegraphics[height=4.2 cm]{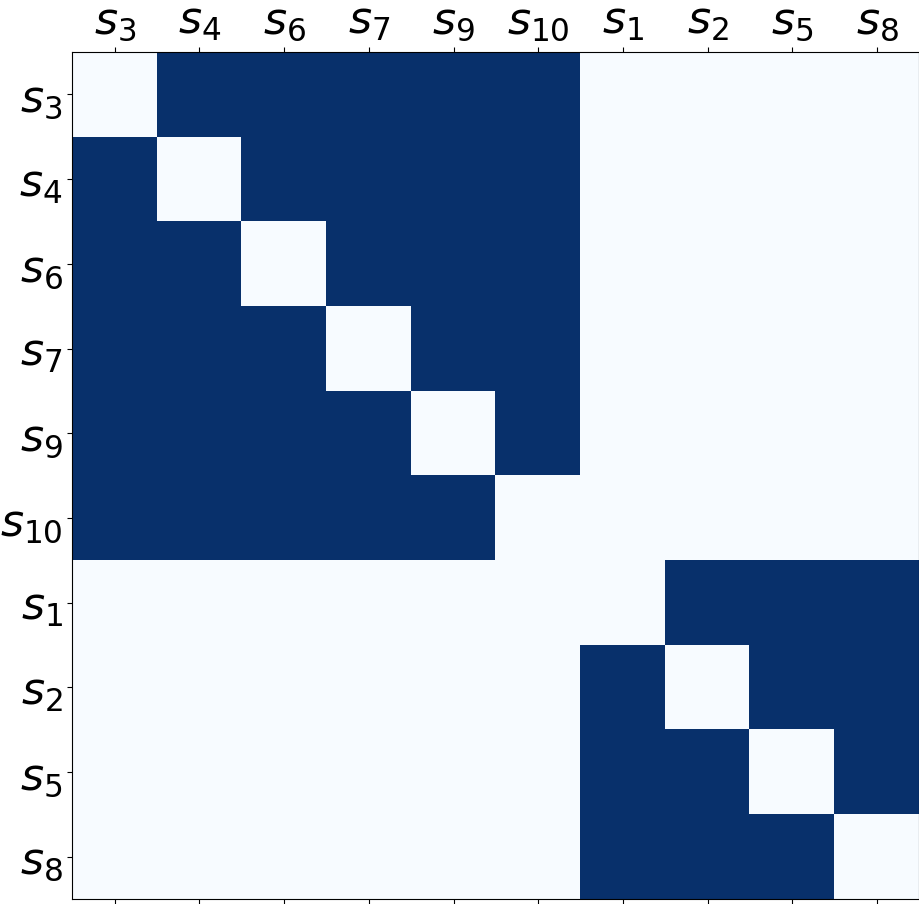}}\hfill
    \subfloat{\includegraphics[height=4.2 cm]{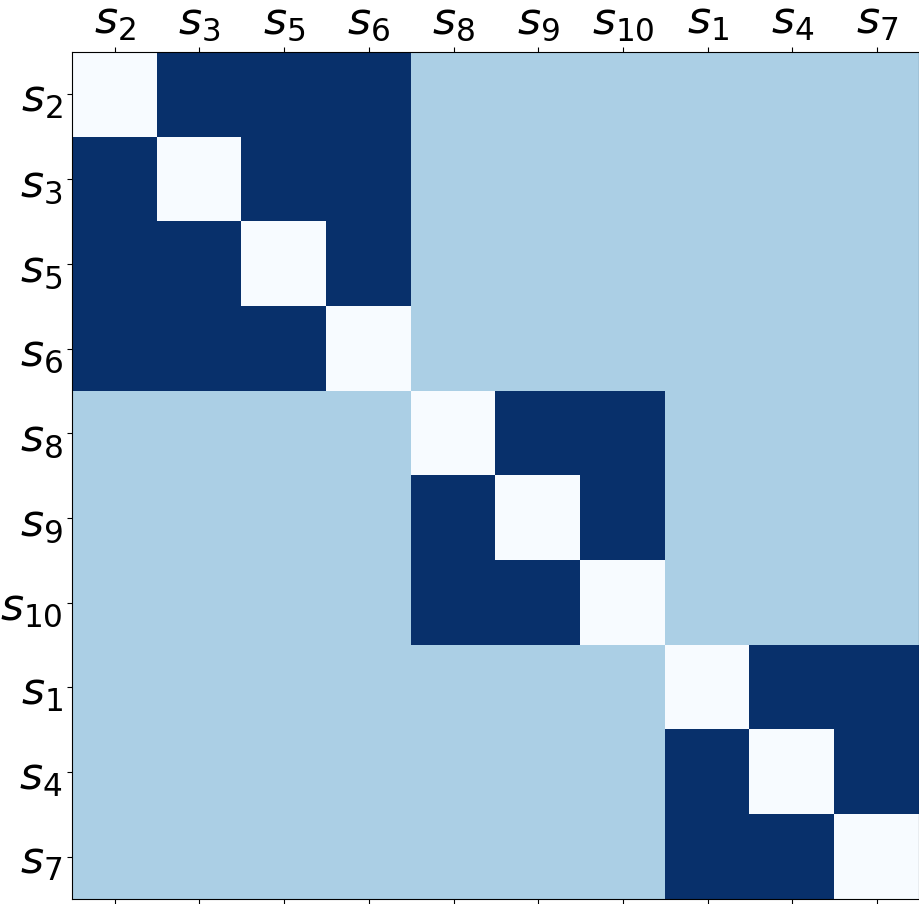}}\hfill
    \subfloat{\includegraphics[height=4.2 cm]{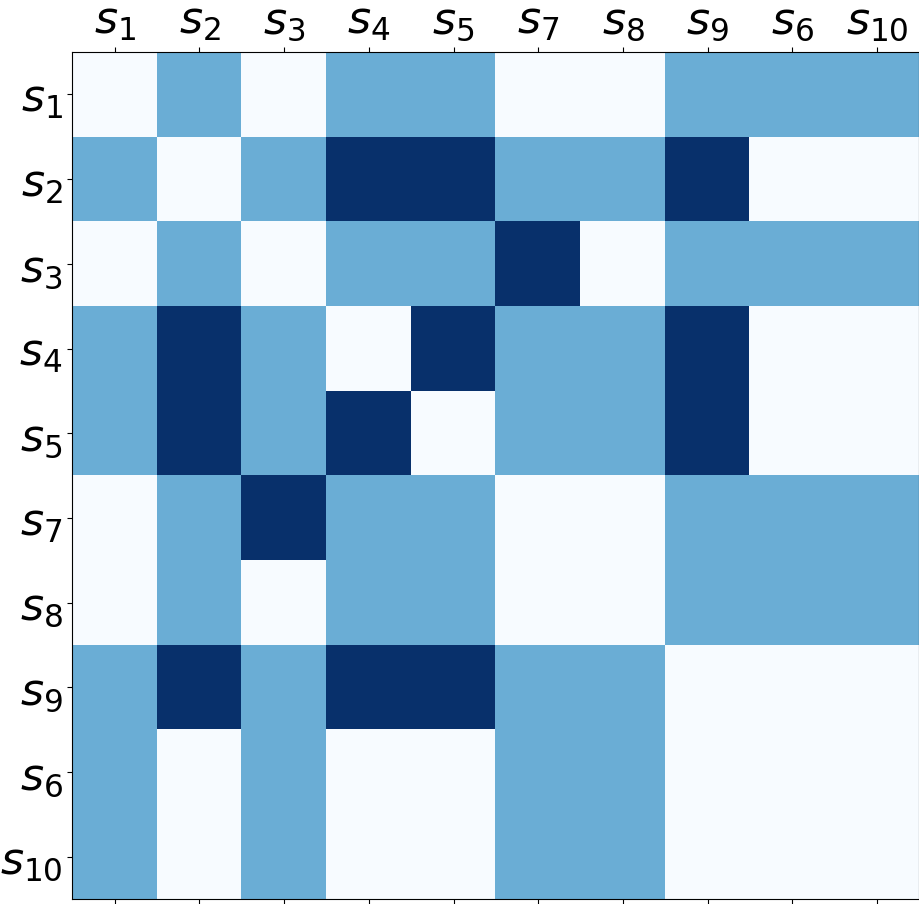}}\\[2ex]
    \subfloat{\includegraphics[height=4.2 cm]{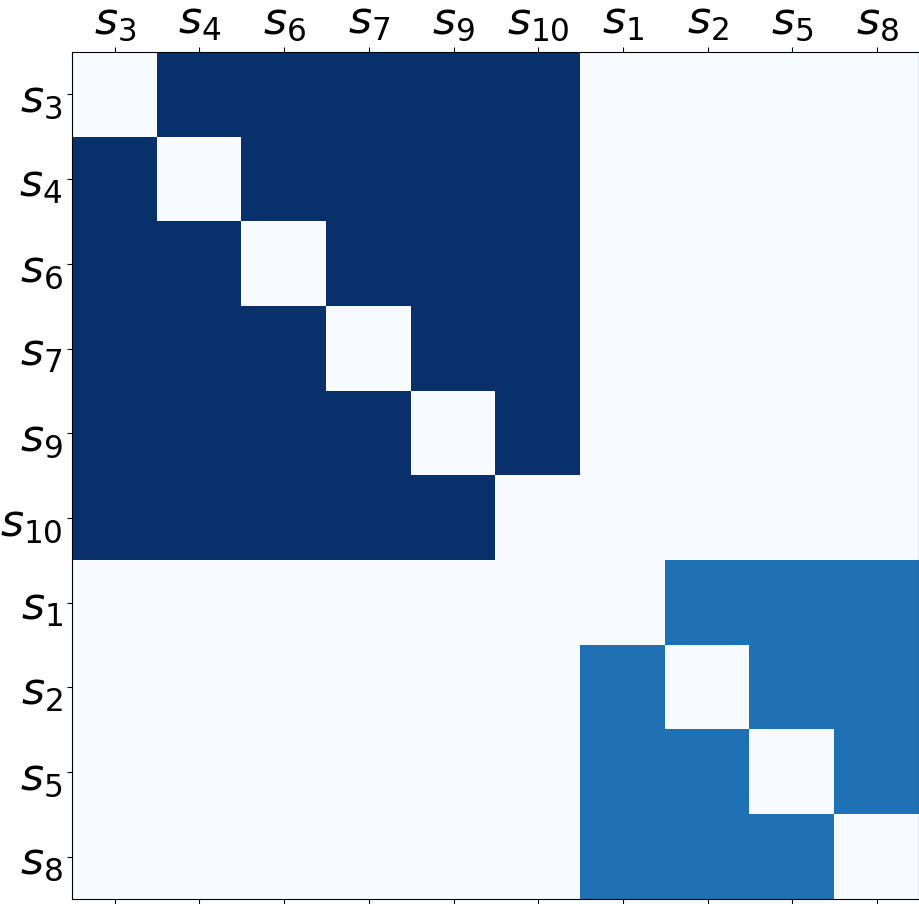}}\hfill
    \subfloat{\includegraphics[height=4.2 cm]{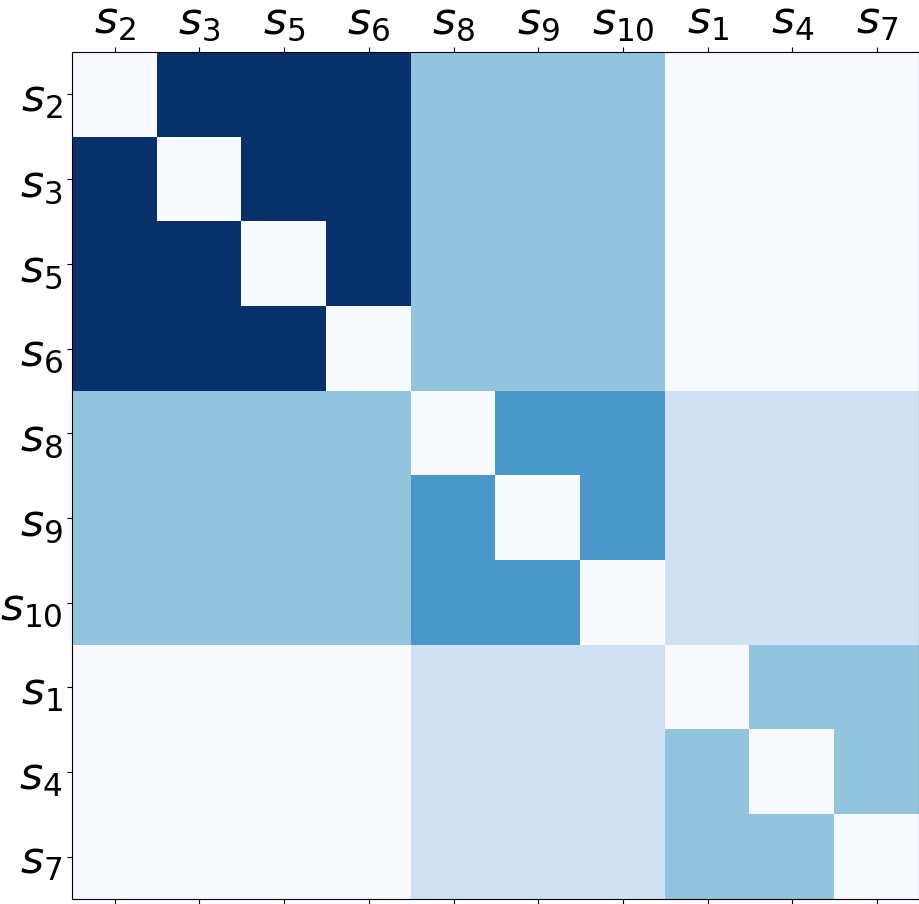}}\hfill
    \subfloat{\includegraphics[height=4.2 cm]{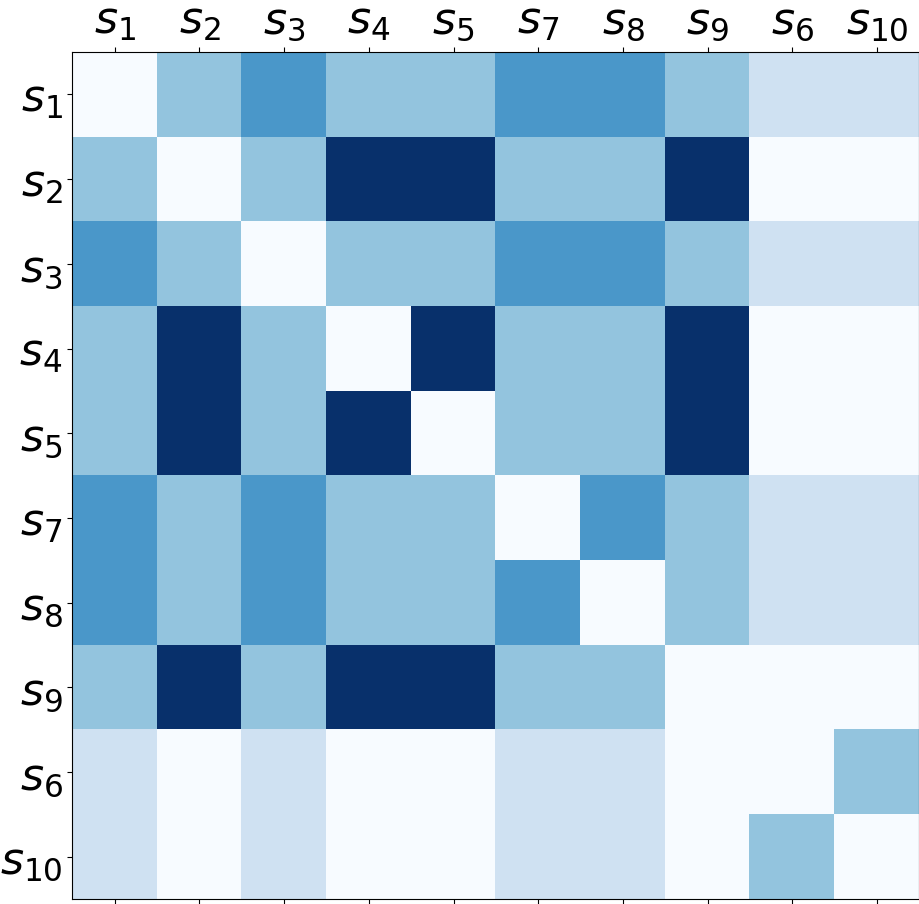}}\\[2ex]
    \subfloat[$p=2$]{\includegraphics[height=4.2 cm]{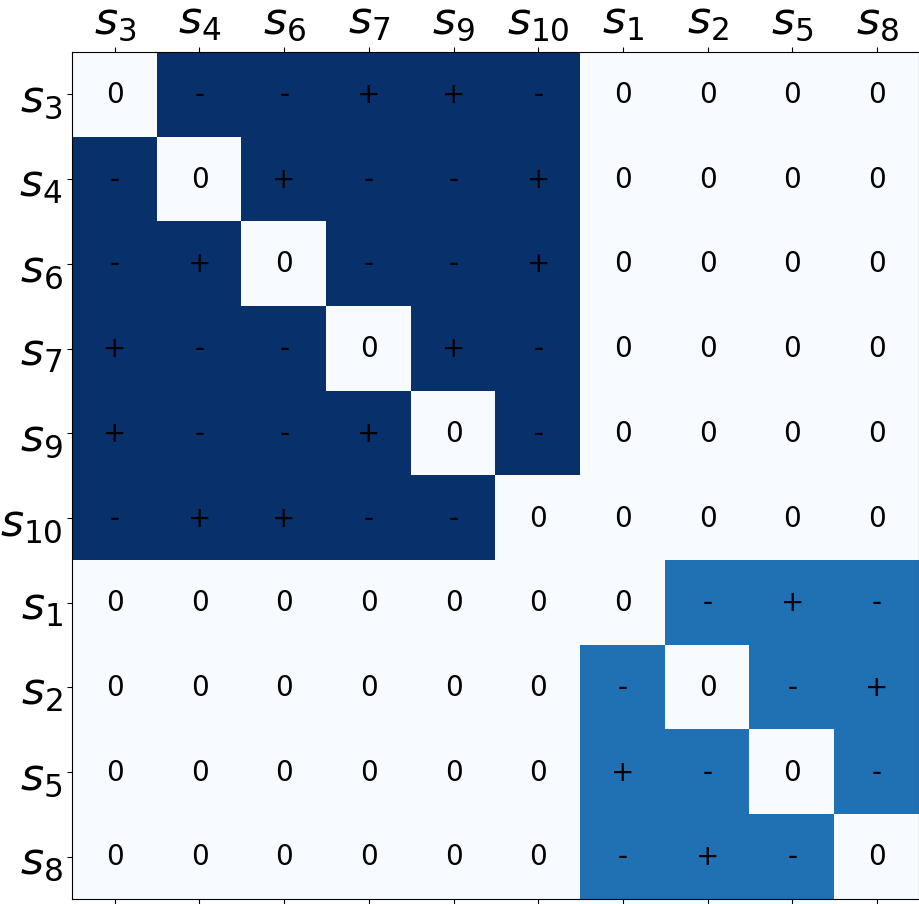}}\hfill
    \subfloat[$p=3$]{\includegraphics[height=4.2 cm]{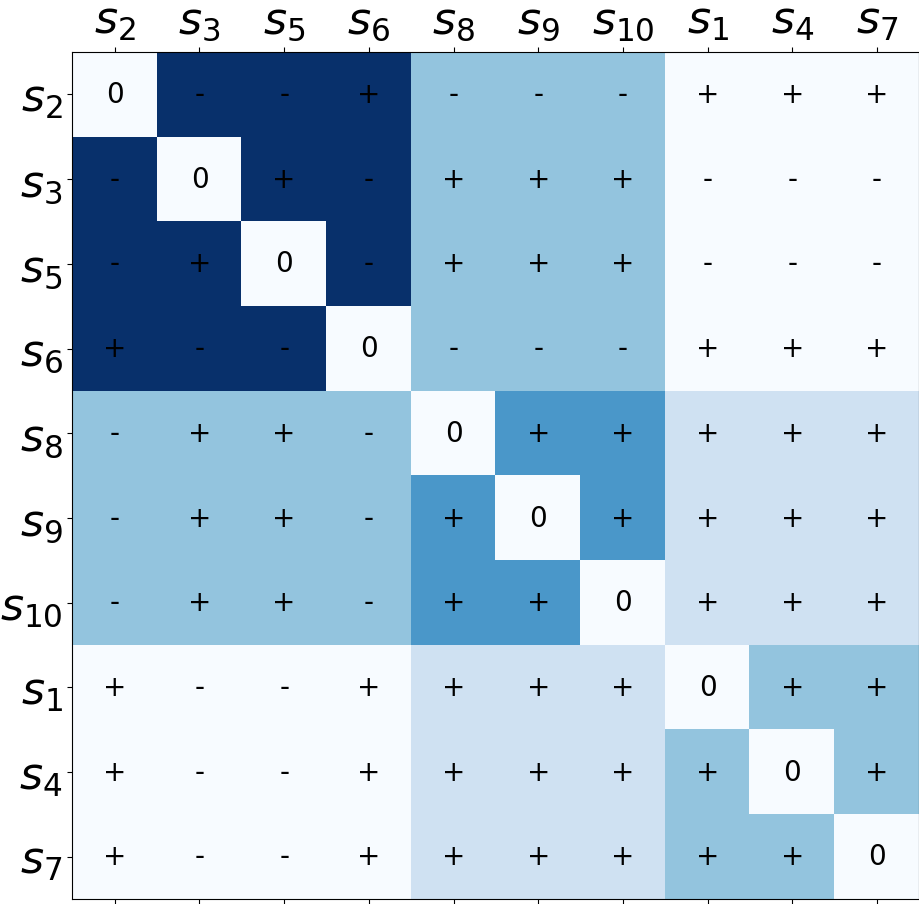}}\hfill
    \subfloat[$p=4$]{\includegraphics[height=4.2 cm]{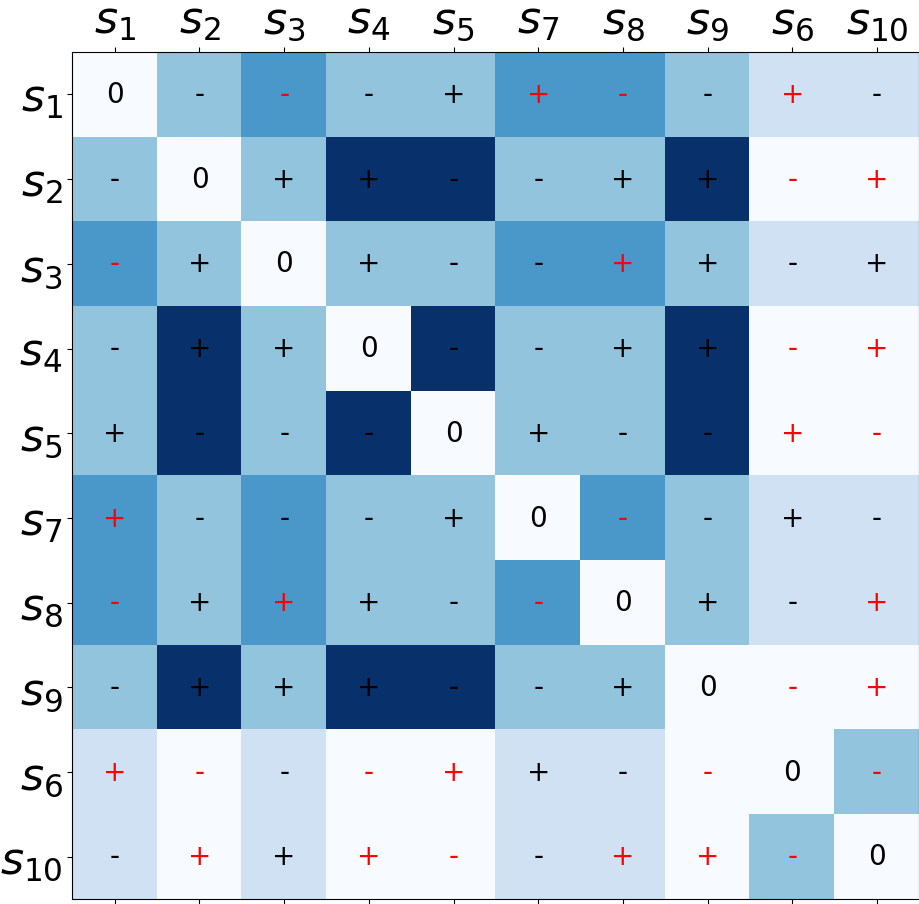}}\\
    \caption{Heat maps of the synaptic matrices of three $N=10$ Hopfield models with different numbers of memorized patterns $p$. The darker the shade of blue, the stronger the coupling (in absolute value). Top row: the three original matrices (absolute values). Middle row: the matrices reconstructed with the information derived from the relevant neurons identified by the mapping entropy. Bottom row: the same matrices of the middle row but with the addition of the signs obtained from the analysis of time correlations $C_{ij}$. The red symbols indicate those signs that differ from the real ones.}
    \label{fig:Smatrix}
\end{figure*}

Let us now investigate more in detail the behaviour of mapping entropy for a varying number of retained neurons in the low-resolution description of the network; in particular, we focus on the specific set of mappings that minimises the mapping entropy at any given value of $n_{cg}$, representing the maximally informative reduced representations that can be designed for the $N=10$ Hopfield network at each degree of resolution. The results of this analysis are presented in Fig.~\ref{fig: Smap_N10_MIN}, and highlight that although, as previously discussed, $S_{map}$ decreases for increasing $n_{cg}$, such decrease is interrupted by some ``flattenings'' that occur between pairs of consecutive values $(\Bar{n}_{cg}, \Bar{n}_{cg} + 1)$. This behavior is more evident for $p=2$ and $p=3$, reduces to a softer change of slope for $p=4$ and vanish at $p=1\ ,5$. Critically, such ``steps'' in $S_{map}$ suggest that adding a neuron to a maximally informative mapping with $n_{cg} = \Bar{n}_{cg}$ retained neurons allows little or no further information gain about the system; the retained spin set minimizing the mapping entropy at fixed $n_{cg} = \Bar{n}_{cg}$ thus stands out from the others. This behaviour is comparable to the one observed by Giulini and coworkers \cite{HGP} for a discrete system of non-interacting spins, where they observed an entire range of $n_{cg}$ values featuring almost the same mapping entropy minimum. In that case, this behaviour could be explained by noting that the mappings minimising $S_{map}$ most frequently included spins whose probability to be in a given state (e.g. $+1$) was appreciably different from that of the others, the latter being prone to be treated as noise. Here we obtained a similar result, despite the crucial difference of dealing with a strongly interacting system.

Based on the observations discussed insofar, one could expect the group of relevant neurons $\{ s_{i_{\nu}}^*\}_{\nu=1,...,\Bar{n}_{cg}}$ highlighted by the step in $S_{map}$ to be characterized by strong couplings among them. We could also argue that the couplings with neurons outside this group should be weaker, which is another argument in favour of the interpretation of the discarded neurons' dynamics as noise. In fact, the presence of a further neuron strongly coupled with the group of retained ones would have shifted the flattening to $n_{cg} = \Bar{n}_{cg} + 1$, and this additional neuron would have been itself a part of the retained group. In light of these considerations, what these results suggest is that the minimization of the mapping entropy highlights which neurons are practically decoupled from the rest of the network.

In order to test the validity of this hypothesis for the set of relevant neurons, we reshuffled the order of the spins inside the synaptic matrix, grouping the retained spins $\{ s_{i_{\nu}}^*\}$ in the first $\Bar{n}_{cg}$ columns/rows; we then calculated then matrix semi-dispersion as:
\begin{eqnarray}\label{eq:semi_dispersion}
    &&\rho =  \frac{\langle \abs{J} \rangle_{s^*s^*} - \langle \abs{J} \rangle_{ss^*}}{\langle \abs{J} \rangle_{s^*s^*} + \langle \abs{J} \rangle_{ss^*}}\\ \nonumber
    &&J = 
    \left(
    \begin{array}{c|c}
        s^*s^* & s^*s\\
        \hline
        s\,\,s^* & s\,\,s
    \end{array}
    \right),
\end{eqnarray}
where $ \langle \abs{J} \rangle_A$ stands for the absolute coupling averaged over the elements of block $A$. $\rho$ quantifies in a single value the properties of the retained neurons we want to investigate, that is, the strength of the couplings within the group of retained spins relative to that of the couplings between retained and discarded ones. If the retained spins interact much more strongly among themselves than with the discarded ones, $\rho > 0$.

To gain further insight, we identify and rank the previously discussed ``steps'' that appear in the plots of $S_{map}$ vs. $H_{S}$, see Fig.~\ref{fig: Smap_N10_MIN}, by calculating the increment of the discrete first derivative of $S_{map}(H_S)$ (something akin to a finite-difference second order derivative):
\begin{equation}\label{eq:Delta}
\begin{split}
    \Delta(n_{cg}) =& \frac{S_{map}(n_{cg} - 1) - S_{map}(n_{cg})}{H_{s}(n_{cg} - 1) - H_{s}(n_{cg})} +\\
    &- \frac{S_{map}(n_{cg}) - S_{map}(n_{cg} + 1)}{H_{s}(n_{cg}) - H_{s}(n_{cg} + 1)}.
\end{split}
\end{equation}

With this definition, we have that for $n_{cg} = \Bar{n}_{cg}$ the quantity $\Delta$ is minimized. Table \ref{tab_semidsp} shows the three mappings with lowest values of $\Delta$ (ordered by $\Delta$ in ascending order) and their related semi-dispersions $\rho$ for the $p=3$ model; the last column reports the maximum possible value of semi-dispersion at fixed $n_{cg}$ for the $p=3$ synaptic matrix.

\begin{table}
\begin{center}
\begin{tabular}{c|c|c|c|c}
\centering
        $n_{cg}$ & Mapping & $\Delta$ & $\rho$ & $\rho_{max}$\\
        \hline
        4 & [$s_2,s_3,s_5,s_6$] & $-0.85$ & $0.50$ & $0.50$\\
        \hline
        7 & [$s_2,s_3,s_5,s_6, s_8, s_9, s_{10}$] & $-0.81$ & $0.30$ & $0.30$\\
        \hline
        2 & [$s_2,s_5$] & $0.14$ & $0.41$ & $0.41$
\end{tabular}
\end{center}
\caption{List of the mappings that minimise the quantity $\Delta$ defined in Eq. \ref{eq:Delta}. $n_{cg}$ is the number  of neurons retained in the mapping; the \textit{Mapping} column lists the spins contained in the low-resolution representation; the last three columns report the values of $\Delta$, the semi-dispersion $\rho$, and the largest value of the latter that can be attained in the system for that particular coarse-graining level.}
\label{tab_semidsp}
\end{table}

The results reported in Tab.~\ref{tab_semidsp} confirm our expectations, and so do those for $p=2$ and $p=4$ (data not shown), but the observed signal weakens for higher numbers of memorized patterns. This happens for two reasons: first, with higher numbers of memories the system moves towards the spin glass phase, and the efficiency of memory retrieval decreases because of the rising internal noise \cite{hopfield1982neural} and due to the growth of non-retrieval states \cite{MBF}. Second, increasing $p$ the synaptic matrix will statistically feature fewer strongly coupled neurons.

\begin{figure*}[t]
    \subfloat[$p=2$, biased patterns]{\includegraphics[height=4.4 cm]{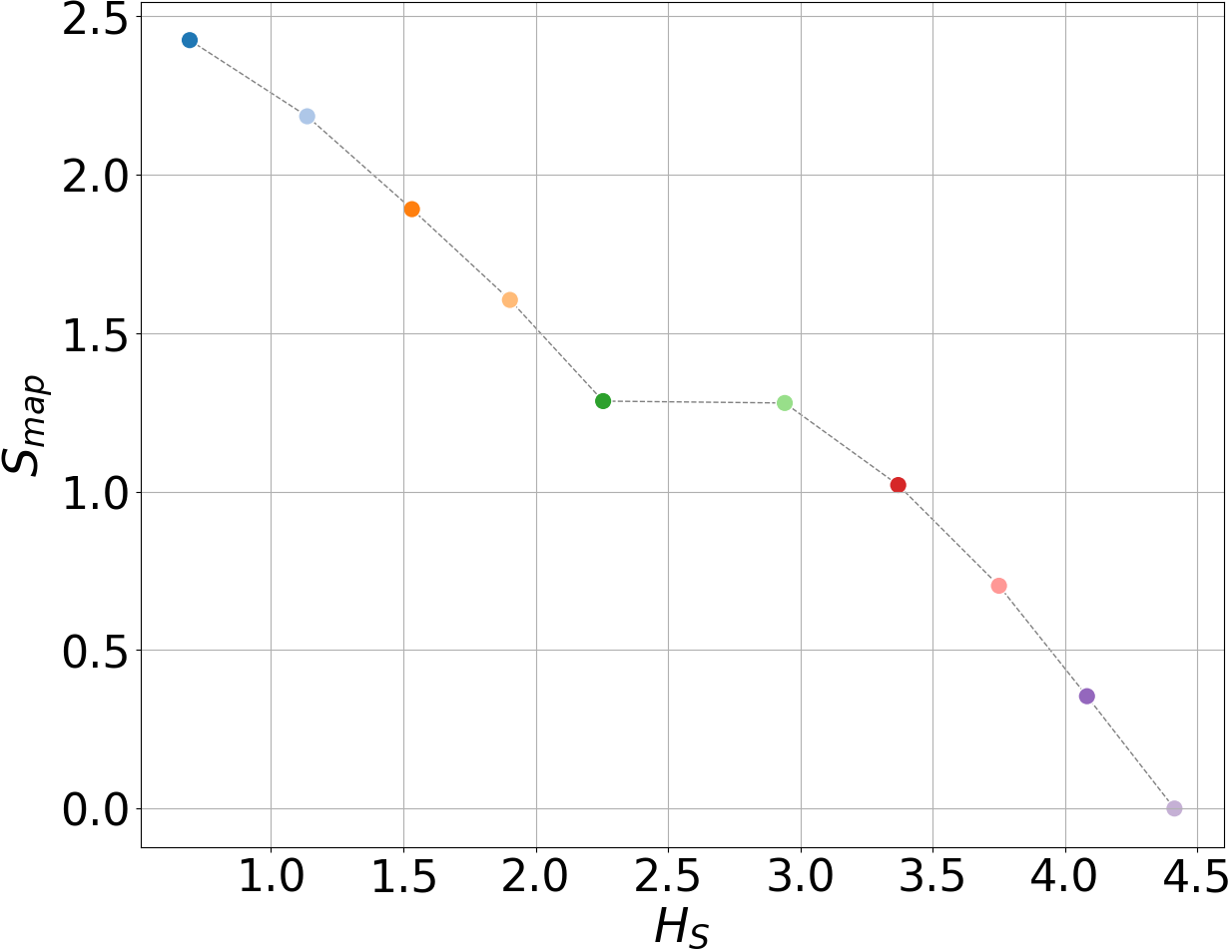} \label{fig:Smap_N10_p2_BIASED}}\hfill
    \subfloat[$p=5$, random patterns]{\includegraphics[height=4.4 cm]{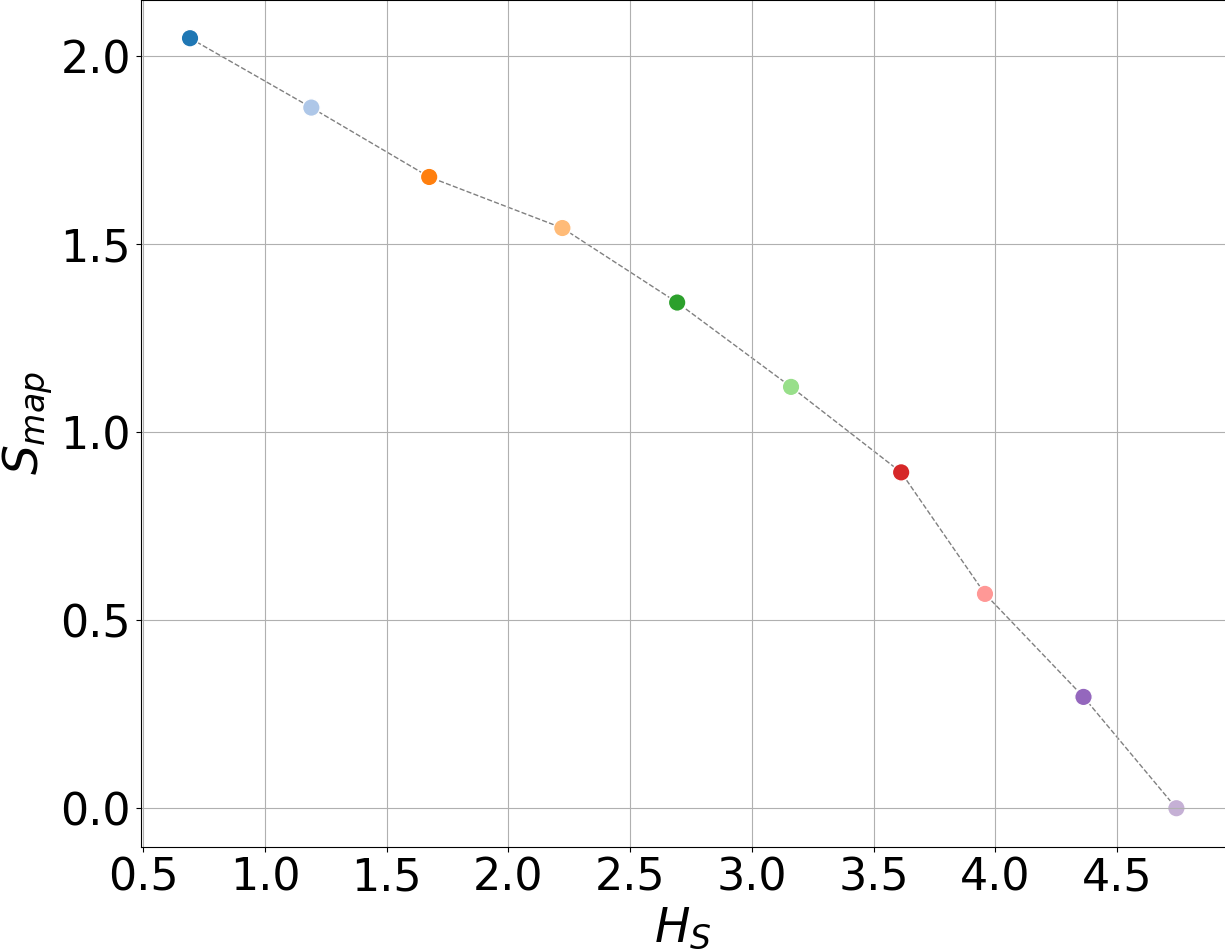} \label{fig:Smap_N10_p5_UNBIASED}}\hfill
    \subfloat[$p=5$, biased patterns]{\includegraphics[height=4.4 cm]{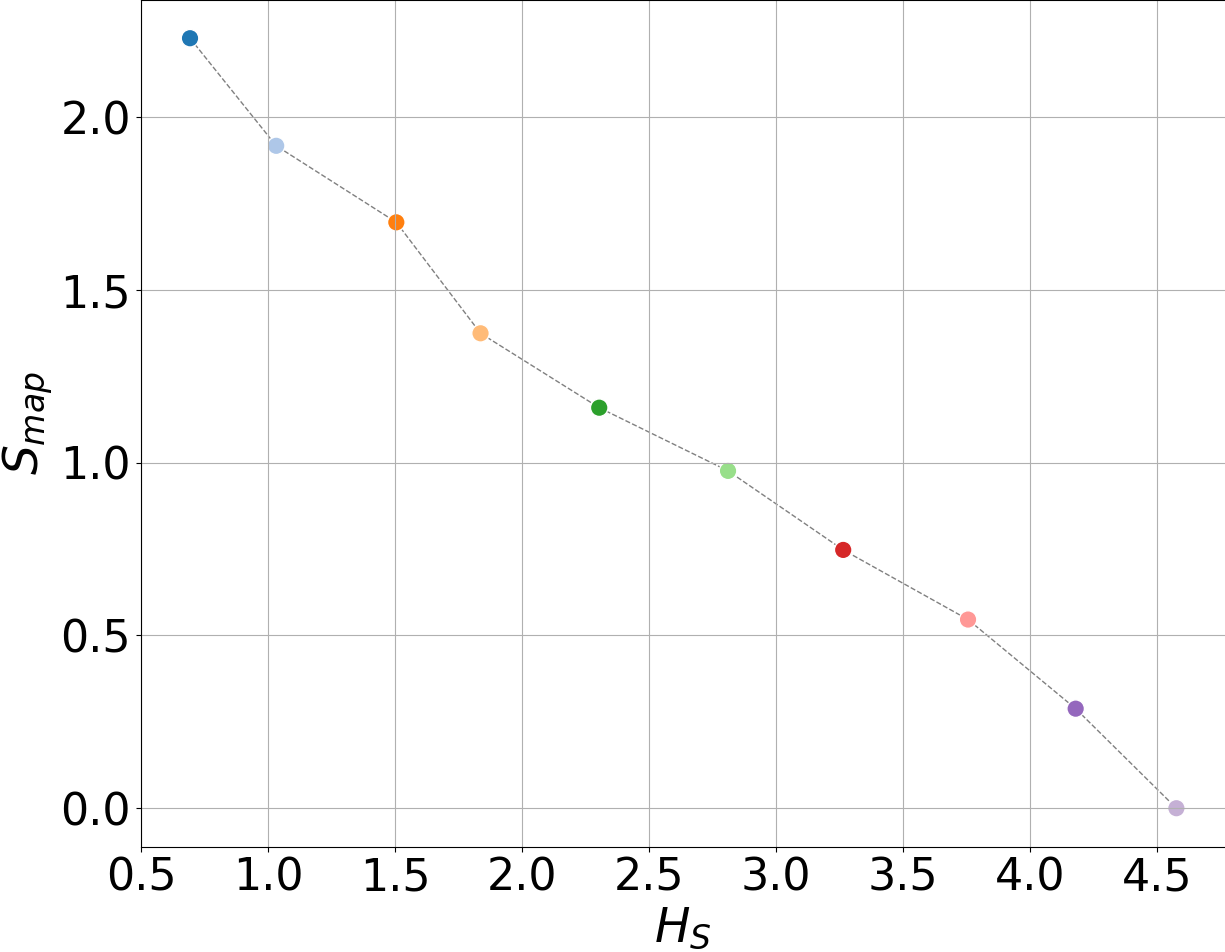} \label{fig:Smap_N10_p5_BIASED}}
    \begin{center}
        \subfloat{\includegraphics[width=4.5 cm]{figures/Legend/N10.png}}
    \end{center}   
    \caption{Minima of the mapping entropy plotted as a function of resolution for every number $n_{cg}$ of retained spins. Panel (a): results from a simulation of an $N=10$, $p=2$ Hopfield model with biased patterns described by Eq. \ref{eq:biased_patterns_p2}. Panel (b): results from the same simulation of an $N=10$, $p=5$ Hopfield model presented in Fig. \ref{fig:MIN_N10_p5}. Panel (c): results from a simulation of an $N=10$, $p=5$ Hopfield model with biased patterns described by Eq. \ref{eq:biased_patterns_p5}.}
\end{figure*}

\subsubsection{Approximate reconstruction of the interaction matrix from the optimal mappings}

The information about the coupling among neurons, which the minimisation of the mapping entropy allowed us to infer, can be leveraged to reconstruct, although approximately, the entire synaptic matrix $J$. To this end, we employ the data pertaining to the first two groups of informative neurons, identified thanks to Eq. \ref{eq:Delta}. More specifically, we first calculate the value of $\Delta(n_{cg})$ for every CG mapping with minimum value of mapping entropy at fixed $n_{cg}$; second, we rank these groups of neurons by increasing value of $\Delta$; third, we focus on the first two groups of this rank. We then make use of the fact that these groups of neurons feature high values of synaptic matrix semi-dispersion, and assume that this is due to strong couplings inside each group and weak couplings with discarded neurons.

In practice, we initialise our reconstructed $S$ matrix as a null matrix, so that all neurons are initially decoupled. We then leverage our knowledge about the matrix semi-dispersion and increase by $1$
\footnote{We chose to increase and decrease the couplings by an arbitrary value since we are not interested in the absolute magnitude of the coupling (the global order of magnitude of the couplings is irrelevant for the dynamics of the Hopfield model \cite{hopfield1982neural}), but rather on the relative magnitude among couplings.}
the value of all couplings within the first group of most informative (high semi-dispersion) neurons. Conversely, we decrease by $1$ all couplings between the aforementioned neurons and the other ones (we recall that high semi-dispersion implies the presence of a group of strongly coupled neurons with week interactions with the others). We repeat the same procedure with the second group of informative neurons. What we get at the end of this heuristic procedure is a reconstructed $S$ matrix with positive and negative couplings. The elements of this matrix are not an estimate of the real couplings of the model, but rather an approximated ranking of their absolute strength. This means that the higher the value given to a ``reconstructed'' coupling, the stronger the real coupling (in absolute value) should be compared to the other ones.

To give an example, we can look at Tab.\ref{tab_semidsp} and try to reconstruct the synaptic matrix of the model in absolute value. The group of neurons corresponding to the lowest value of $\Delta$ is $\{ s_2, s_3, s_5, s_6 \}$. Thus we proceed by increasing by $1$ all couplings between these neurons and decreasing by $-1$ all couplings between them and the remaining group $\{ s_1, s_4, s_7, s_8, s_9, s_{10} \}$. Then we consider the second lowest value of $\Delta$ and its related mapping $\{ s_2, s_3, s_5, s_6, s_8, s_9, s_{10} \}$ and perform the same modifications to the matrix as before. What we get at the end is the following matrix:
\begin{equation}
    S=  
        \begin{pmatrix}
        $0$ & $-2$ & $-2$ & $0$ & $-2$ & $-2$ & $0$ & $-1$ & $-1$ & $-1$\\
        $-2$ & $0$ & $2$ & $-2$ & $2$ & $2$ & $-2$ & $0$ & $0$ & $0$\\
        $-2$ & $2$ & $0$ & $-2$ & $2$ & $2$ & $-2$ & $0$ & $0$ & $0$\\
        $0$ & $-2$ & $-2$ & $0$ & $-2$ & $-2$ & $0$ & $-1$ & $-1$ & $-1$\\
        $-2$ & $2$ & $2$ & $-2$ & $0$ & $2$ & $-2$ & $0$ & $0$ & $0$\\
        $-2$ & $2$ & $2$ & $-2$ & $2$ & $0$ & $-2$ & $0$ & $0$ & $0$\\
        $0$ & $-2$ & $-2$ & $0$ & $-2$ & $-2$ & $0$ & $-1$ & $-1$ & $-1$\\
        $-1$ & $0$ & $0$ & $-1$ & $0$ & $0$ & $-1$ & $0$ & $1$ & $1$\\
        $-1$ & $0$ & $0$ & $-1$ & $0$ & $0$ & $-1$ & $1$ & $0$ & $1$\\
        $-1$ & $0$ & $0$ & $-1$ & $0$ & $0$ & $-1$ & $1$ & $1$ & $0$
        \label{S_biased_XXX}
        \end{pmatrix}
        .
\end{equation}

Fig. \ref{fig:Smatrix} shows some of the reconstructed $S$ matrices (second row) we obtained with this method compared with the real ones (first row). The last row of the figure shows the results of a tentative derivation of the coupling signs, which would complete the synaptic matrix reconstruction. In order to determine the sign of a coupling $J_{ij}$, we compute the correlation $C_{ij} \equiv \langle s_i s_j \rangle$ between the corresponding spin pair, averaging over the whole configuration sample. The coupling $J_{ij}$ is attributed the sign of the associated correlation; in contrast, the coupling is set to $0$ if $|C_{ij}| < 0.1 C^\star$, where $C^\star \equiv \sup_{ij} |C_{ij}|$ is the largest absolute value of the correlation between all spin pairs. The results mirror the original signs with the exception of a few errors for the $p=4$ model.

Another way to leverage the information given by $S_{map}$ is to identify possible biases introduced in the definition of the model. If we take, for example, the $N=10$, $p=2$ Hopfield model with patterns given by Eq. \ref{eq:biased_patterns_p2}, the mapping entropy curve in Fig.~\ref{fig:Smap_N10_p2_BIASED} highlights that increasing from $5$ to $6$ the number of retained neurons does not correspond to an appreciable increase in the information content of the reduced representation (that is, $S_{map}$ remains practically the same). Notably, the particular value $n_{cg} = 5$ is a consequence of the fact that the first five spins have identical states in the memory patterns.

We then apply the same process to a $p=5$ Hopfield model; we thus impose the values of the first $m$ spins of each one of the five patterns so that they are equal or opposite. Consider for example the following case:
\begin{equation}
\begin{tabular}{c c c c c c c c c c c}
    $p_1$: & \textbf{-1} & \textbf{-1} & \textbf{1} & \textbf{1} & -1 & -1 & 1 & -1 & 1 & -1,\\
    $p_2$: & \textbf{-1} & \textbf{-1} & \textbf{1} & \textbf{1} & 1 & 1 & 1 & 1 & 1 & 1,\\
    $p_3$: & \textbf{1} & \textbf{1} & \textbf{-1} & \textbf{-1} & -1 & -1 & 1 & -1 & 1 & -1,\\
    $p_4$: & \textbf{1} & \textbf{1} & \textbf{-1} & \textbf{-1} & 1 & 1 & 1 & -1 & -1 & 1,\\
    $p_5$: & \textbf{-1} & \textbf{-1} & \textbf{1} & \textbf{1} & 1 & -1 & -1 & 1 & -1 & -1
\end{tabular}\label{eq:biased_patterns_p5}
\end{equation}

Here we imposed that the first $m = 4$ spins have either equal memories in patterns $p_1$, $p_2$, $p_5$ and opposite ones in $p_3$, $p_4$. In this way, the couplings $J_{ij}$ for $i\neq j \in \{1,2,3,4\}$ will all have maximum absolute value equal to $p/N$. Figures \ref{fig:Smap_N10_p5_UNBIASED} and \ref{fig:Smap_N10_p5_BIASED} show a comparison between the minima of the mapping entropy for simulations with random and biased patterns, respectively. As we previously argued, in the absence of a bias the model with $p=5$ doesn't present any clear decrease of the information loss rate, as quantified by the quantity $\Delta$ in Eq.~\ref{eq:Delta}, while increasing $H_S$, see Fig.~\ref{fig: Smap_N10_MIN}; on the contrary, in Fig.~\ref{fig:Smap_N10_p5_UNBIASED} we can see that this rate actually increases, that is, the $S_{map}$-vs.-$H_S$ curve becomes steeper. This does not happen in the case of biased patterns: Fig.~\ref{fig:Smap_N10_p5_BIASED} shows that, after the $S_{map}$ minimum related to $n_{cg}=4$, the information loss rate actually decreases slightly. The group of relevant neurons highlighted by the mapping entropy is thus composed by four spins and these correspond exactly to the first four biased spins $\{ s_1, s_2, s_3, s_4 \}$.

\subsection{Decimation regimes for a N=100 Hopfield model}
\label{sec:n100network}

\begin{figure*}
    \subfloat[$p=4$]{\includegraphics[height=6.8 cm]{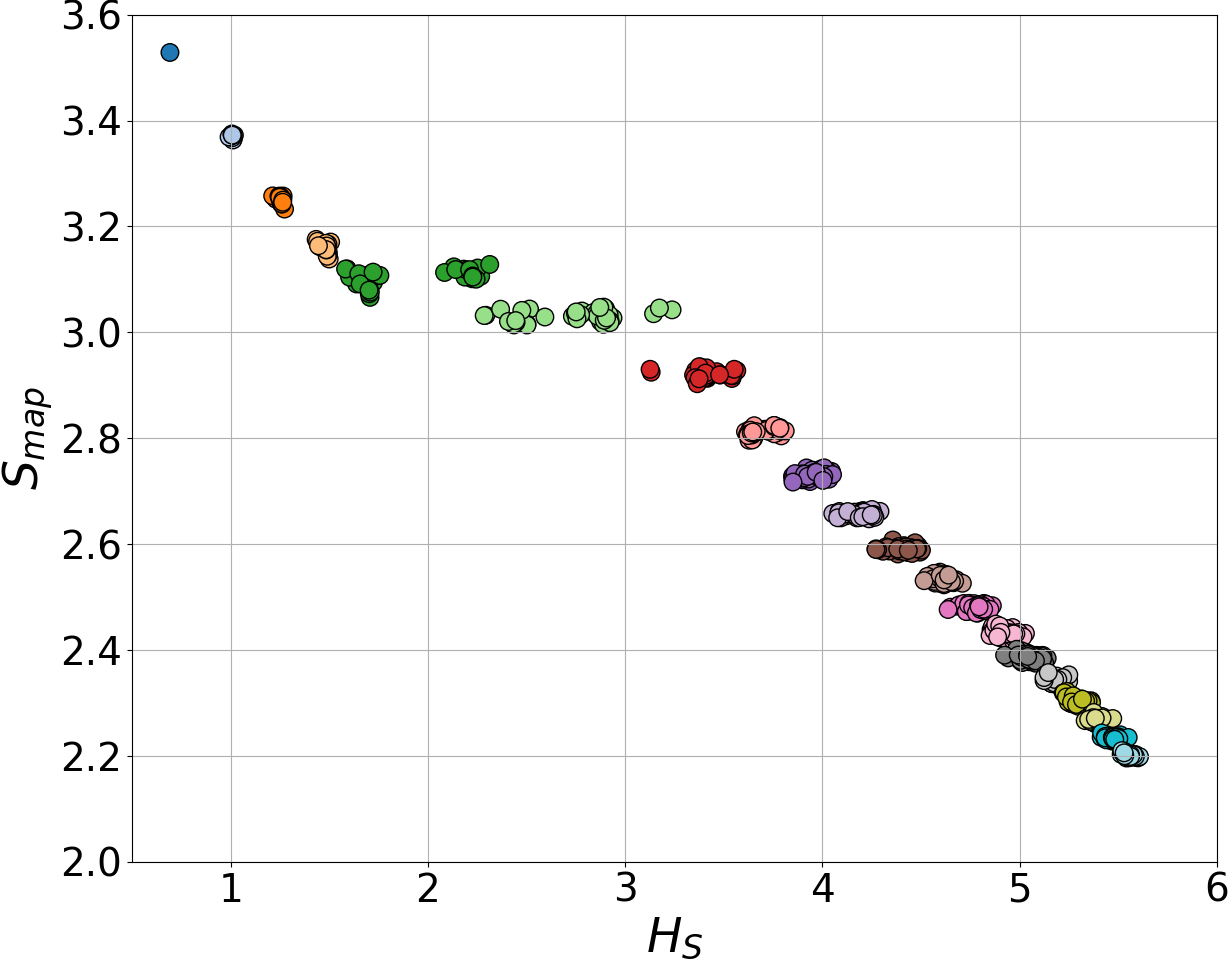}}\hfill
    \subfloat[$p=5$]{\includegraphics[height=6.8 cm]{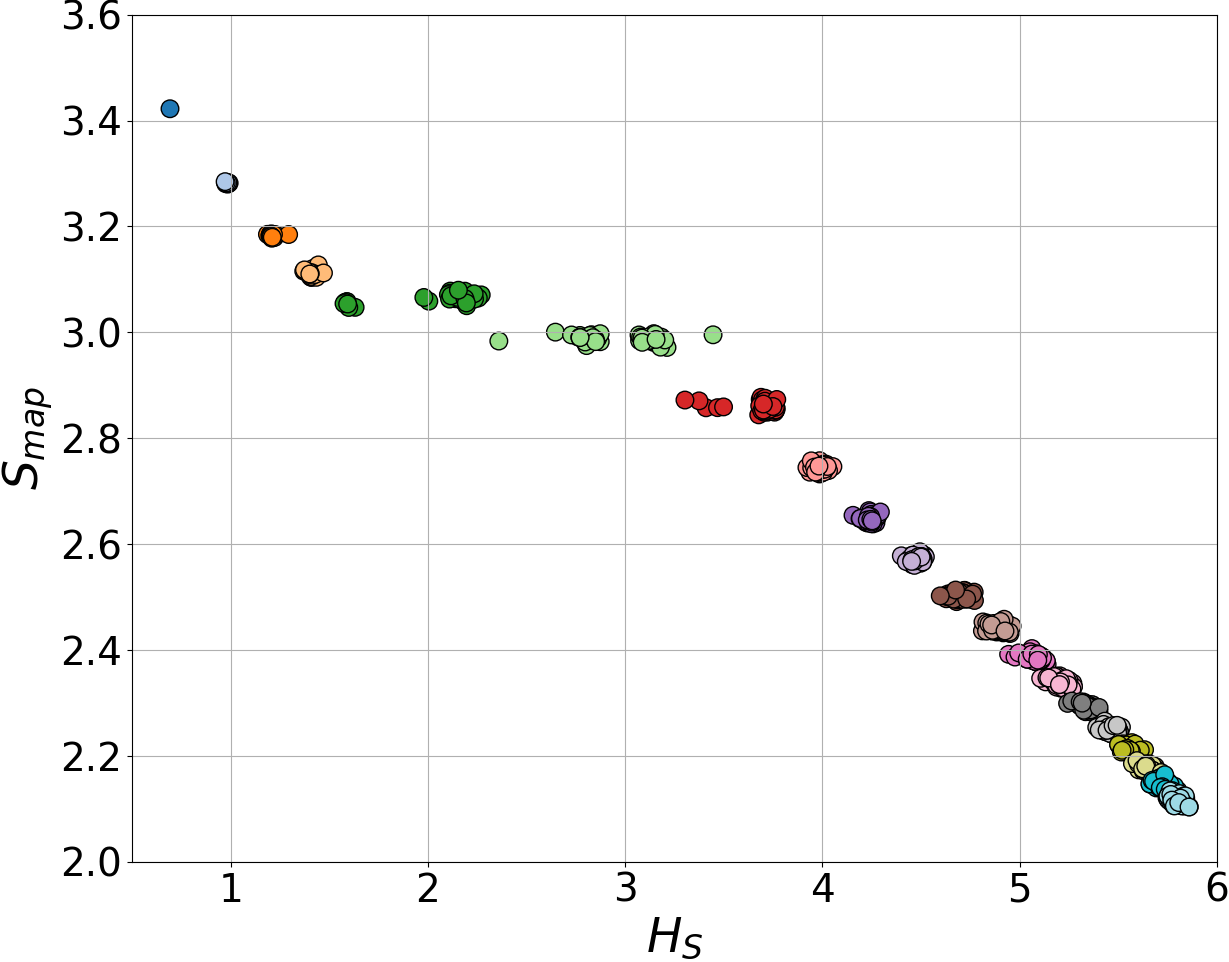}}
    \begin{center}
        \subfloat{\includegraphics[width=8 cm]{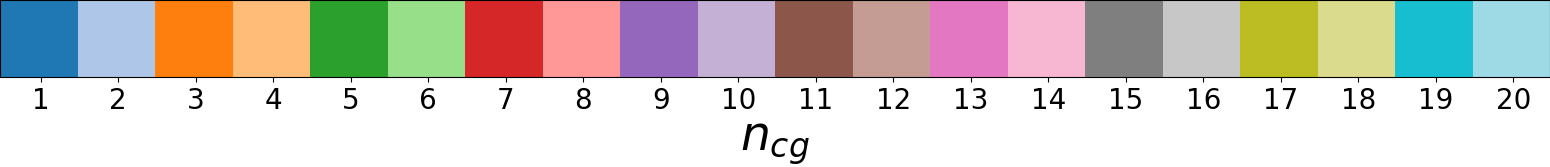}}
    \end{center}
    \caption{Mapping entropy minima for a Hopfield model with $N=100$ neurons, for $n_{cg} = 1,\ \cdots \ 20$ plotted as a function of resolution for $p = 4$ (panel a) and $p = 5$ (panel b). The simulated Hopfield model had $N=100$ neurons.}
    \label{fig:Smap_N100_p45}
\end{figure*}

In this section we push forward the analysis carried out in the previous paragraphs, through the application of the MEOW approach to a Hopfield network constituted by a greater number of spins. We first report on the strategies implemented to overcome the challenges that a larger network presents to the identification of the mapping entropy minima; then, we show that this more complex system displays particularly interesting characteristics, regarding in particular the curve of mapping entropy minima as a function of the resolution of the reduced representations; finally, we inspect the features of the interactions between the groups of spins that constitute the optimal mappings, and how they change as the resolution of the simplified model is increased.

The larger the size of a neural network, the richer and more complex its behaviour, which is at least in part the reason behind the emergent properties of our brain \cite{MBF}. Dealing with a finite but large number of neurons $N$, however, makes the network investigation through the MEOW approach more difficult; in fact, the number of possible mappings that we have to take into account when performing decimation grows like $2^N$, preventing us from carrying out an exhaustive enumeration of all possible coarse-grained representations of the system. Nonetheless, resuming the discussion in the previous section, what we are interested in is not an extensive analysis of the whole mapping space, but rather the properties of the mappings that minimise the mapping entropy as a function of resolution.

To this end, we implemented the mapping entropy optimisation workflow relying on a simulated annealing (SA) minimisation strategy \cite{ref_62_JCTC, ref_63_JCTC}, as it was originally done by Giulini \emph{et al.} in Ref.~\cite{JCTC}. We performed simulations of an $N=100$ Hopfield network at $T=0.2$ for different values of $p \in $ \{2, 3, ..., 10\}; these sets of parameters were chosen to fall within the retrieval phase of the infinite-size model, see Sec. \ref{subsec:Hopfield_model}. We then looked for the maximally informative low-resolution representations of the system, i.e. the ones that minimise the mapping entropy at fixed $n_{cg} \in$ \{1, 2, ..., 20\}.

\subsubsection{Analysis of the least mapping entropy curve as a function of the resolution of reduced representations}

The plots for a subset of the different results obtained, in particular those that pertain to the lowest values of mapping entropy for $p = 4,\ 5$, are shown in Fig.~\ref{fig:Smap_N100_p45}.

For $2\leq p \leq 8$, starting from the coarser representations, the information loss decreases with an approximately constant rate until it reaches a sort of ``inflection region''. Here, the steepness of the $S_{map}\ vs.\ H_S$ curve decreases considerably before it grows again, and the mapping entropy continues its descent. The inflection region includes mappings relative to two to four values of $n_{cg}$, and shows a smearing of these mappings on a broad interval of resolutions. We thus observe a resolution gap between the two regions where the mapping entropy decreases at a constant rate.

\begin{figure*}
    \subfloat[$p=4$]{\includegraphics[height=6.5 cm]{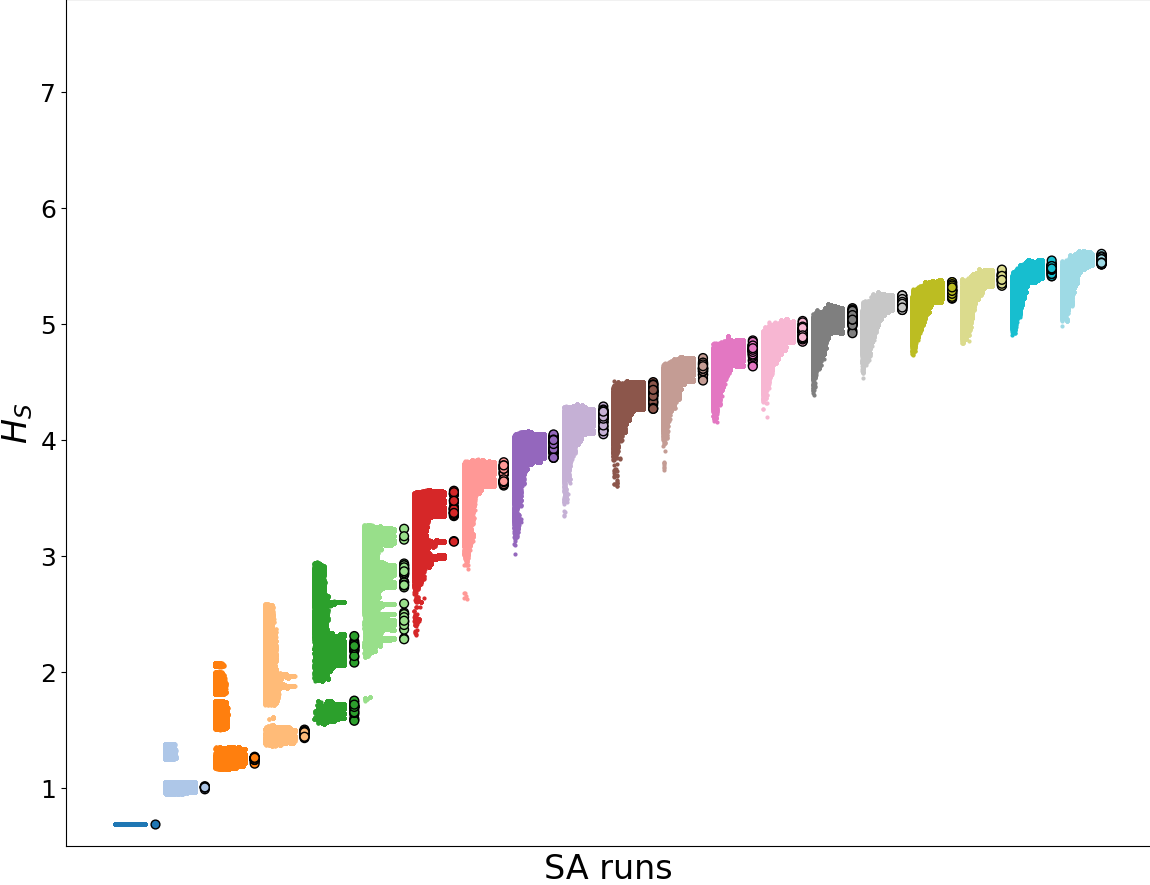}}\hfill
    \subfloat[$p=5$]{\includegraphics[height=6.5 cm]{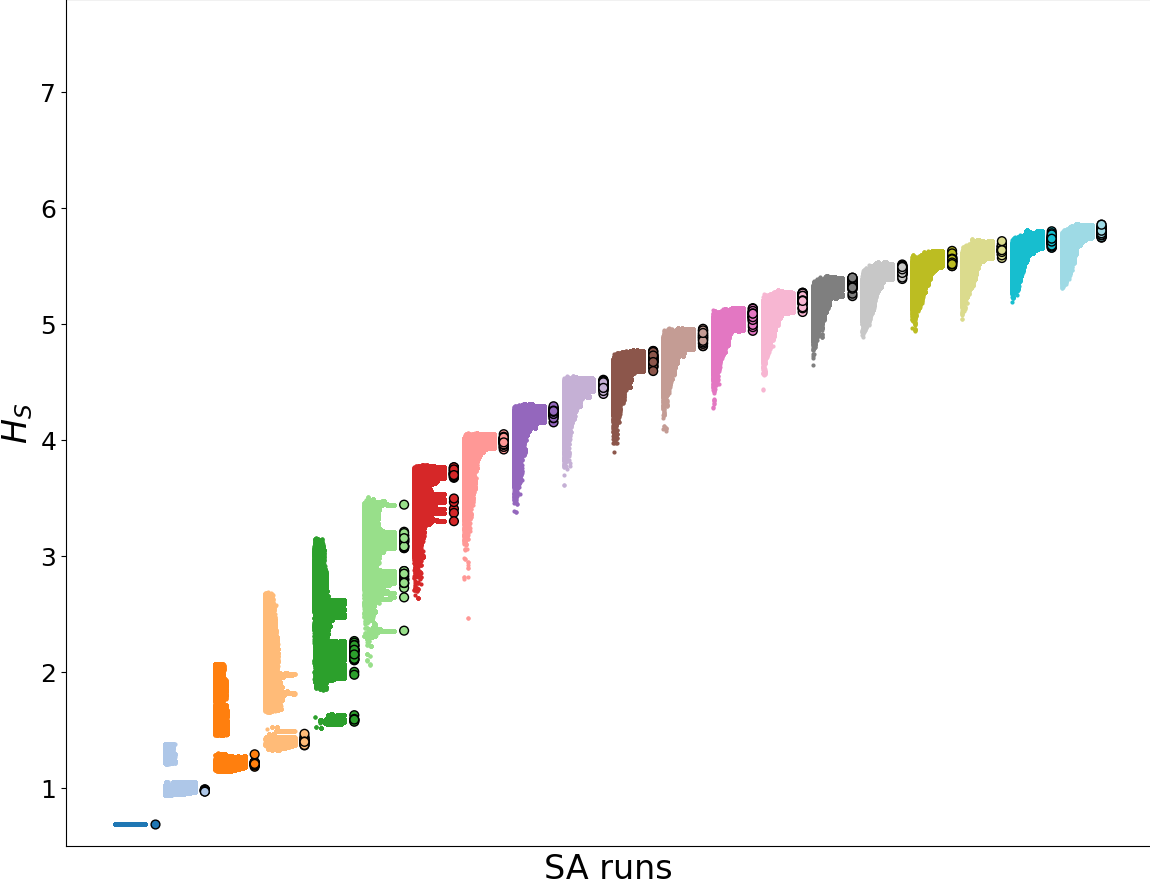}}
    \begin{center}
        \subfloat{\includegraphics[width=8 cm]{figures/Legend/N20.png}}
    \end{center}
    \caption{Resolutions of the reduced representations visited during SA runs at different values of $n_{cg}$ and $p = 4$ (panel a) and $p = 5$ (panel b). The various resolutions explored in one SA process at given $n_{cg}$ and $p$ have the same color, and the ones corresponding to the $S_{map}$ minima found by the algorithm are indicated by black circles. The results for the different runs at fixed $p$ are plotted side by side to make it easier to compare the resolutions at which information loss is minimized for varying numbers of retained neurons.}
    \label{fig:SA_N100_p45}
\end{figure*}

To rationalise these observations we focus on the role played by resolution; more specifically, we exploit the SA algorithm by keeping track of all the $H_S$ values of the mappings visited during the minimization procedure. The results of this analysis for $p=4$ and $p=5$ are shown in Fig.~\ref{fig:SA_N100_p45}.

For values of $n_{cg}$ smaller than the ones that constitute the inflection region ($n_{cg} < 5$ for $p=4,\ 5$), the decimated representations that minimize the mapping entropy are also associated with low values of the resolution; in particular, in the course of the optimisation, the tentative mappings typically have larger values of $H_{S}$ than those that are, eventually, selected for their low $S_{map}$. This behaviour is consistent with the analysis carried out on the $N=10$ model, for which the mapping entropy minimum at fixed $n_{cg}$ had the lowest resolution because of the linear relationship between the two measures, see Eq.~\ref{Eq:Smap-Resolution}.

This picture is challenged by the optimal mappings retaining a number of neurons located after the inflection region ($n_{cg} > 7$ for $p=4,\ 5$); these attain \emph{high} resolution values, see Fig. \ref{fig:SA_N100_p45}, that is, the SA minimization of the mapping entropy converges towards low-compressed, higher-resolution decimated representations. Mappings that decrease the resolution $H_S$, in fact, contribute to lowering $S_{map}$ if and only if the term $\sum_\phi p(\phi) \text{log}(\Omega(\Psi_\phi))$ (Eq. \ref{Eq:Smap-Resolution}) does not increase. Here, instead, we have that low values of mapping entropy correspond to a high $H_S$ counterbalanced by a decrease in the configuration space term, which is in turn due to the limited number of resolved decimated configurations \cite{HGP}.

Finally, the inflection region, where we see a smearing of the resolution values for fixed $n_{cg}$ and almost constant $S_{map}$, corresponds to a changeover situation in which informative mappings cover almost the entire range of the available values of resolution. This means that both high coding cost and highly compressed representations can be found, that lose the same amount of information.

\begin{figure*}
    \subfloat[$p=2$]{\includegraphics[height=4.25 cm]{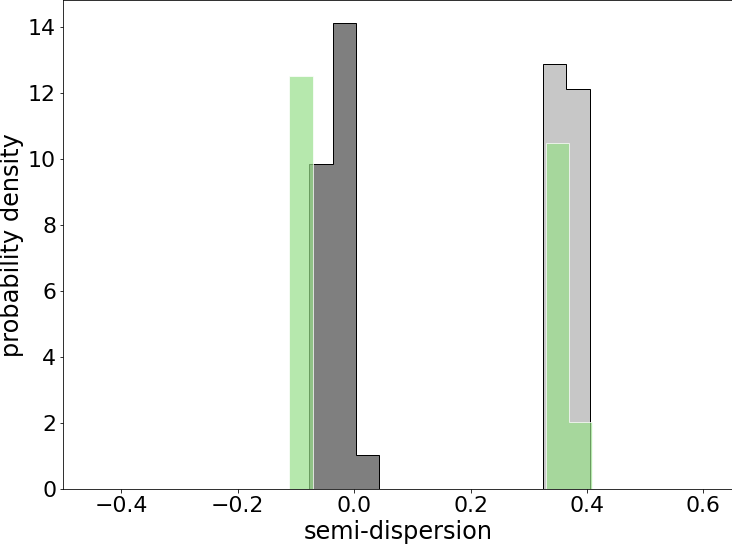}}\hfill
    \subfloat[$p=3$]{\includegraphics[height=4.25 cm]{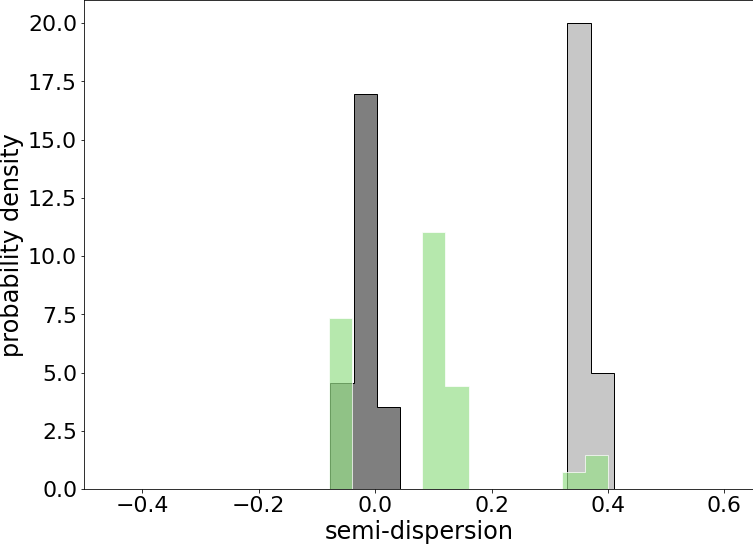}}\hfill
    \subfloat[$p=4$]{\includegraphics[height=4.25 cm]{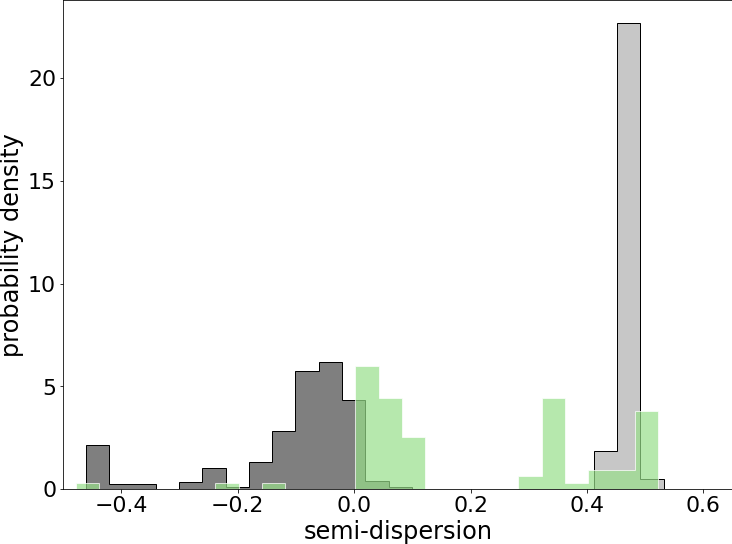}}\\
    \subfloat[$p=5$]{\includegraphics[height=4.25 cm]{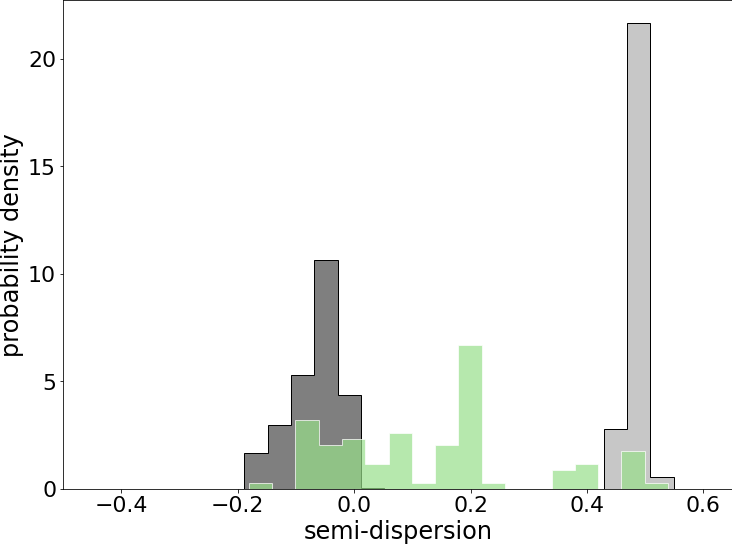}}\hfill
    \subfloat[$p=6$]{\includegraphics[height=4.25 cm]{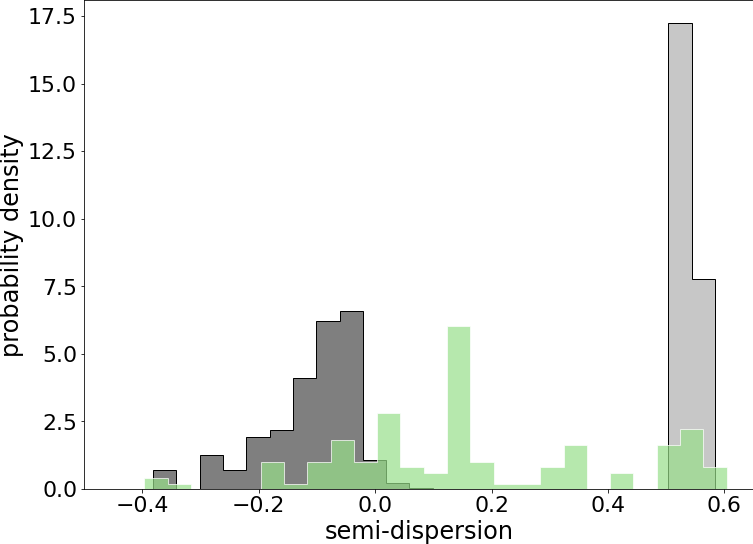}}\hfill
    \subfloat[$p=7$]{\includegraphics[height=4.25 cm]{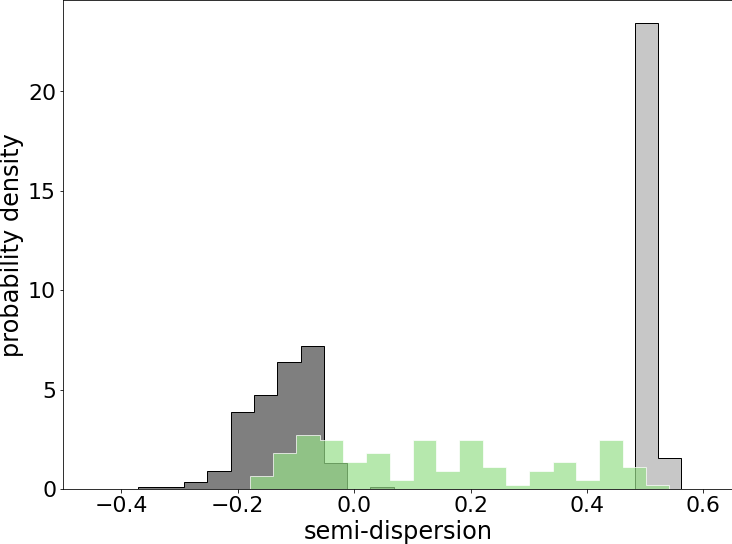}}\\
    \subfloat[$p=8$]{\includegraphics[height=4.25 cm]{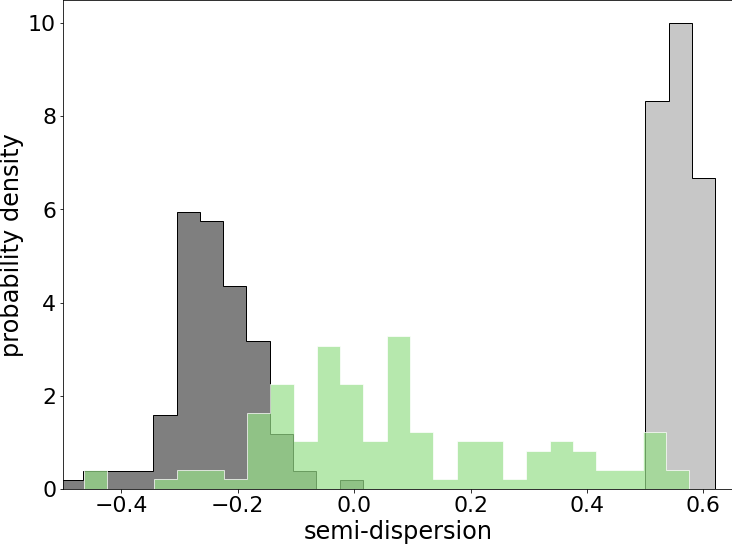}}\hfill
    \subfloat[$p=9$]{\includegraphics[height=4.25 cm]{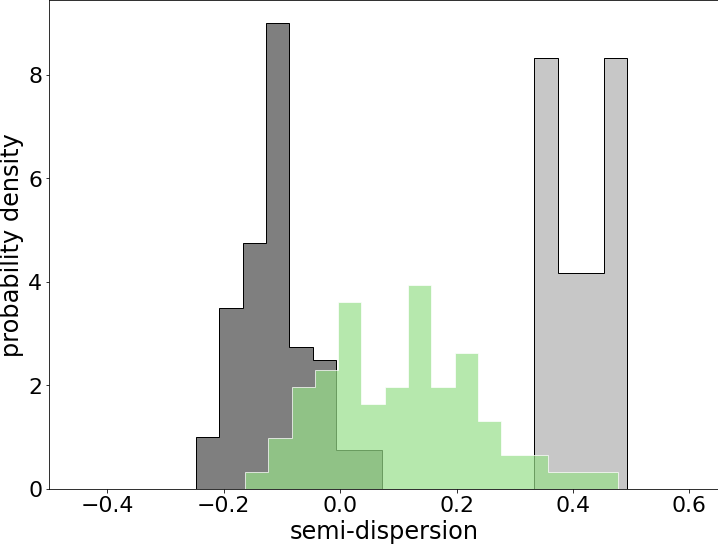}}\hfill
    \subfloat[$p=10$]{\includegraphics[height=4.25 cm]{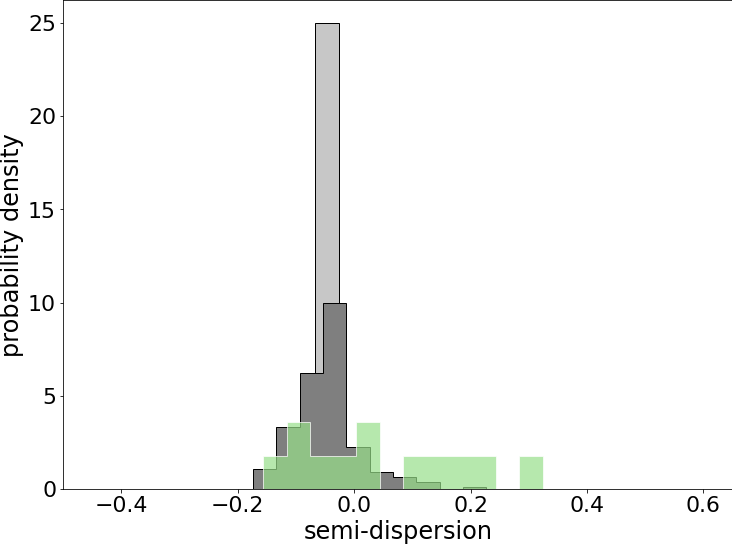}}\\
    \caption{Probability densities of the semi-dispersion for the three different resolution regimes. In light gray: $n_{cg}$'s lower than the inflection region. In dark gray: $n_{cg}$'s higher than the inflection region. In green: $n_{cg}$'s that correspond to the inflection region. Each panel corresponds to a different value of $p$, ranging from $2$ to $10$.}
    \label{fig:main_semi_disp}
\end{figure*}
 
Another informative way of reading the $S_{map}-H_S$ graphs is to inspect the low $S_{map}$ mappings starting from those with low $n_{cg}$ (on the left of the inflection region) and studying their properties as the resolution is increased. As we pointed out for the model with $N=10$, the presence of a sudden decrease in the information loss rate suggests that increasing the detail of our reduced representations does not provide any significant gain in information about the reference system. The minimisation of the mapping entropy is then calling our attention to these optimal mappings, which we can investigate by looking at their values of synaptic matrix semi-dispersion, see Eq. \ref{eq:semi_dispersion}.
 
Fig. \ref{fig:main_semi_disp} displays, for each value of $p$ under examination, the semi-dispersion distributions of the mappings that minimize the mapping entropy. We intentionally distinguished between mappings relative to resolutions lower than, included in, or higher than the inflection region. This highlights once again a gap that separates mappings before and after this region: low-resolution mappings (in light grey) present high values of semi-dispersion, while the latter is low or even negative for high-resolution mappings (in dark grey). The mappings that form the inflection region (shown in green in the plots) take instead intermediate semi-dispersion values, signaling a transition between these two regimes.

We thus have two distinct regimes of decimated representations, corresponding to different values of the resolution as well as the semi-dispersion, separated by a third, crossover interval. Mappings tend to fall either in one regime or the other depending on their number of retained neurons, with the exception of a few specific values of $n_{cg}$ that constitute the intermediate phase where the distribution is wider and covers a large range of semi-dispersion values.

The first regime includes all the mappings with lower numbers of retained neurons: in this case, the mapping entropy minimization favours reduced representations with high values of semi-dispersion and high compression, which means that the most informative way of describing the reference model is to divide it into almost disconnected groups. Every highly compressed informative mapping constitutes a block of strongly interacting neurons that are effectively decoupled from the rest of the network, the dynamics of the latter being seen as an effective noise. The more neurons we retain in our description, the fewer strongly coupled spins are present and the more their dynamics will depend on the discarded neurons, until this strategy eventually becomes inadequate.

The second regime shows us an alternative strategy, which includes all the mappings with higher numbers of retained neurons: when their number becomes too large for a neat decoupling of their group from the rest, the most informative description of the model features null or negative values of semi-dispersion; this indicates that the optimal representation involves weakly interacting neurons, which are often tightly coupled with the discarded ones. When these external interactions are strong, the statistical behaviour of the discarded spins is tightly bound to that of the retained ones; hence, high-resolution, informative mappings are seemingly able to describe the whole network state with the greatest detail, enabling the distinction between the different retrieved patterns.

The third class of mappings to discuss is that of the inflection region, which covers almost entirely the range of semi-dispersion separating the two aforementioned regimes. In this case, both high- and low-resolution descriptions, retaining strongly and weakly coupled groups of retained neurons respectively, can be equally informative low-dimensional representations of the reference system.

The behaviour discussed insofar occurs for all values of $p \leq 9$. For $p = 10$, instead, we have a substantial collapse of all semi-dispersion distributions onto the same range, which indicates the onset of the spin glass phase and the loss of a well-defined crossover region between compressed and low-compression mappings.

\begin{figure*}[th]
    \subfloat[$p=4$]{\includegraphics[height=6.8 cm]{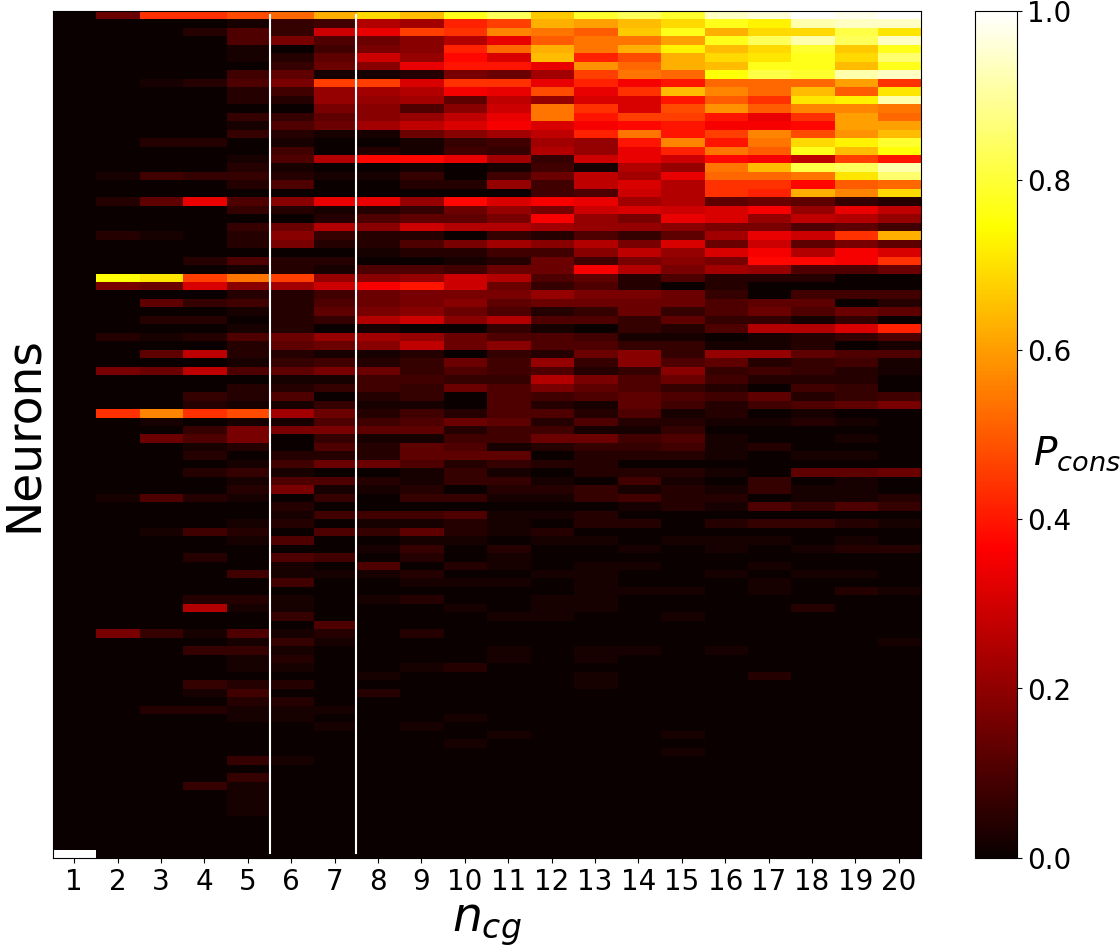}}\hfill
    \subfloat[$p=5$]{\includegraphics[height=6.8 cm]{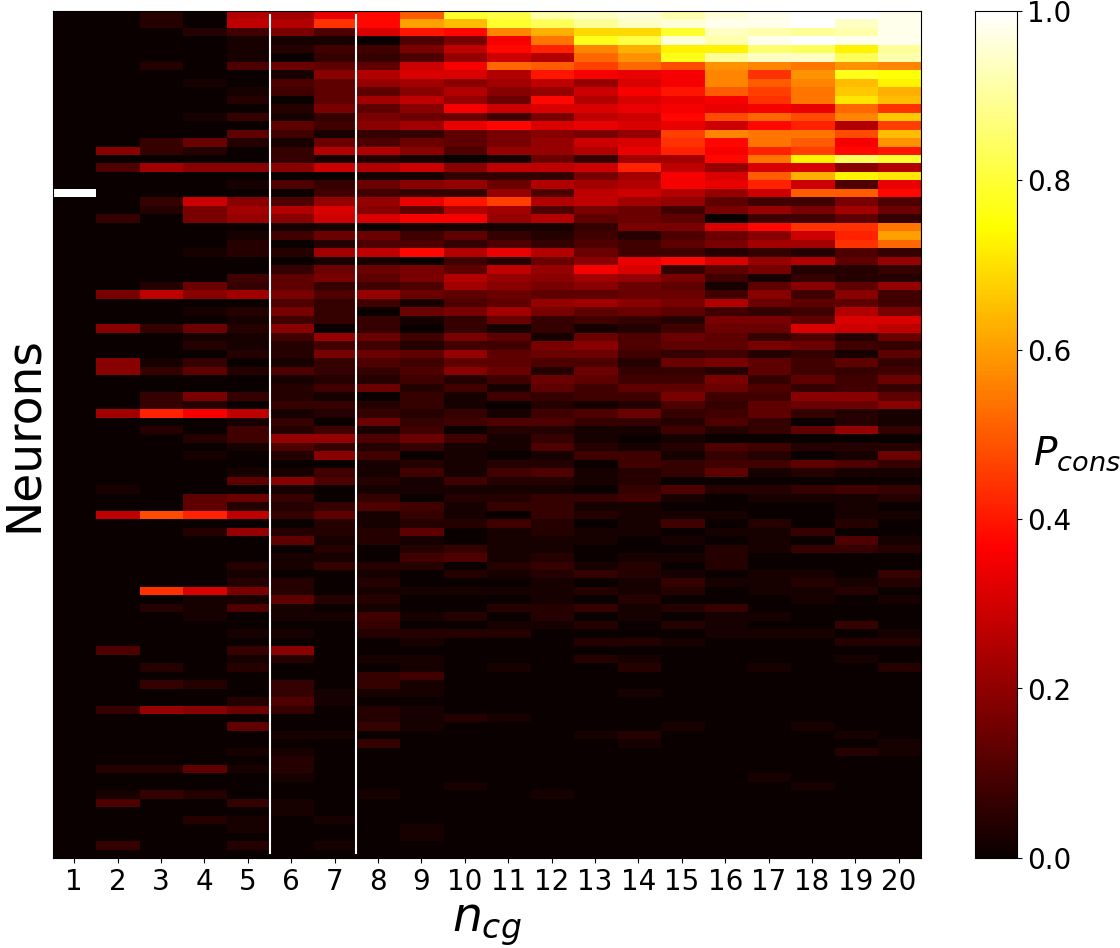}}\\
    \begin{minipage}[c]{0.4\textwidth}
        \subfloat[$p=6$]{\includegraphics[height=6.8 cm]{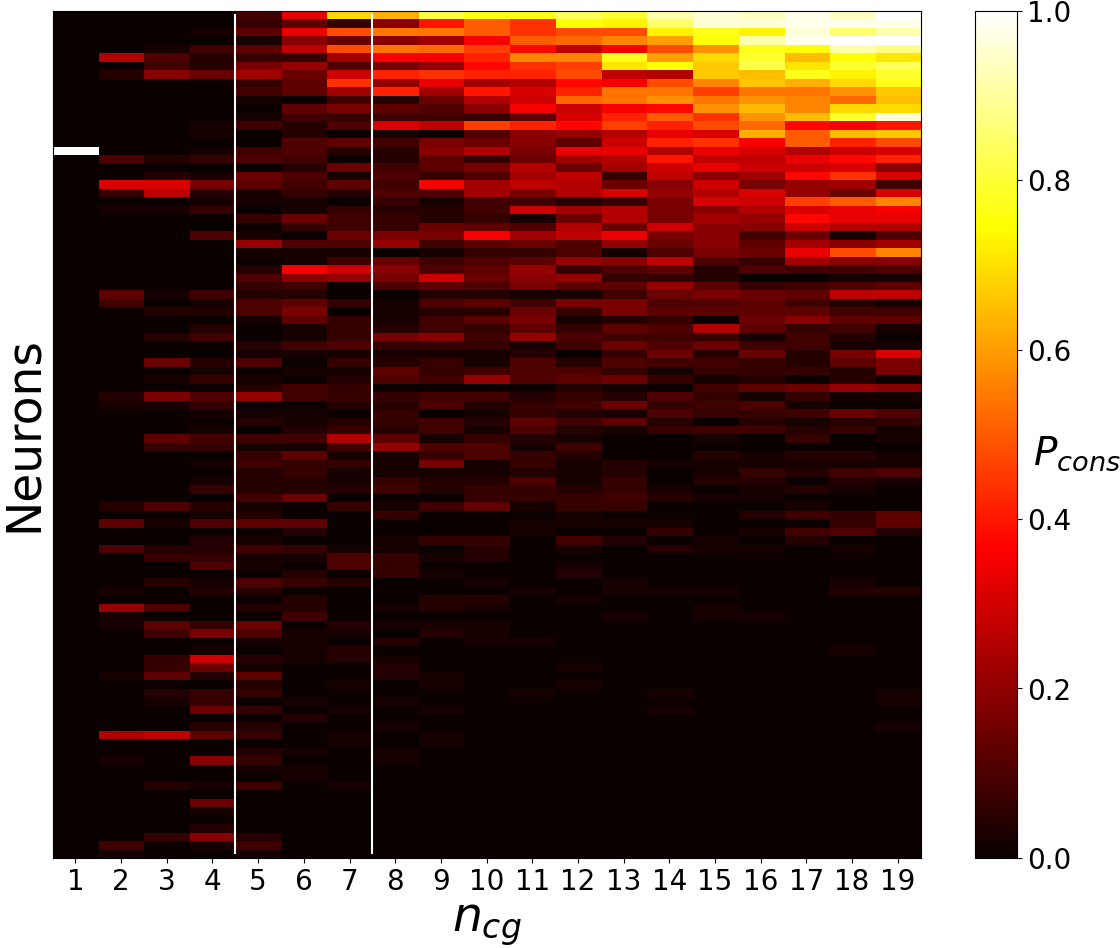}}\hfill
    \end{minipage}\hfill
    \begin{minipage}[c]{0.45\textwidth}
        \caption{Heat maps of the probability $P_{cons}$ that a given neuron is retained in an optimal mapping, plotted as a function of the total number of retained neurons $n_{cg}$ for a model with $N=100$ and $p = 4,\ 5,\ 6$. Each row corresponds to a specific neuron $i$; neurons on the $y$ axis are sorted in descending order of the mean probability $P_{cons}$ to be retained in the high-resolution regime. The latter corresponds to the interval of $n_{cg}$ located of the right of the rightmost white vertical line; correspondingly, the low-resolution and inflection regimes correspond to the first and middle intervals, respectively.}
        \label{fig:prob_cons}
    \end{minipage}
\end{figure*}

\subsubsection{Properties of the neurons most frequently retained in the optimal mappings as function of the resolution}

Lastly, we focus on the probability $P_{cons}$ for an individual neuron to be retained in an optimal mapping. $P_{cons}$ is a function of $n_{cg}$, and it is simply obtained as the empirical occurrence of a given neuron in the pool of least-$S_{map}$ mappings obtained by MEOW. Fig. \ref{fig:prob_cons} illustrates this probability as a function of the number of retained neurons $n_{cg}$. To organise the data, we computed the average $P_{cons}$ for each neuron over the entire high-resolution regime, and sorted them accordingly in decreasing order along the $y$ axis. In the graphs, two vertical lines mark the three regimes previously discussed: it is quite evident that the neurons showing high conservation probabilities in the two main regimes (high and low semi-dispersion) are rarely the same. This is consistent with our understanding of the various $S_{map}$ optimisation strategies: in fact, if a neuron is retained in an optimal mapping ascribed to the high-resolution regime, it would likely present relatively strong couplings with a broad number of neurons. On the other hand, a neuron typically retained in a low-resolution regime mapping will present very strong couplings only with few other neurons, and it is likely to be almost decoupled from the rest. In light of this, we can rationalise why the groups of neurons with high conservation probability in the two sectors relative to these regimes tend to be distinct.

This analysis thus illustrates that the answer to the search for the most informative neurons of a network will critically depend on the chosen resolution scale, that is, the value of $n_{cg}$ in relation to the size of the configuration sample.

\section{Conclusions}\label{sec:conclusions}

In this work we carried out a study of the Hopfield model through the lenses of information-theoretic measures, expanding on the work done by Giulini and coworkers \cite{HGP} and Roudi and coworkers \cite{On_sampling_and_modeling_Marsili, Marsili_1}. Specifically, we investigated the properties of low-resolution representations of specific realisations of the model making use of the mapping entropy optimization workflow (MEOW) and an analysis based on resolution, a measure of the level of detail inherent in a dataset of coarse representations of a system's states. Applying these tools to the configurations sampled in a number of simulations of different Hopfield networks, we observed that the low-resolution mappings with the lowest values of mapping entropy reveal information about intrinsic properties of the reference system.

We first addressed the study of a Hopfield model with $N=10$ neurons. Focusing on the behaviour of the mapping entropy minima at varying resolution, we observed the presence of sudden decreases of the information loss rate that highlighted specific groups of neurons; these proved to have strong interactions among themselves and weak couplings with the discarded ones. As a simple, quantitative measure of this partitioning into groups we employed the semi-dispersion of the average strength of the synaptic matrix elements, which governs the interaction between neurons in the various groups; when looking at the mappings that minimise the mapping entropy in correspondence of these decrease rate variations, we observe that the semi-dispersion is maximized. This is suggestive of the fact that mapping entropy minima are in correspondence of mappings that retain strongly interacting neurons, while the reminder is weakly coupled with the first group as well as within itself. This result is in line with the one discussed in \cite{HGP} for a discrete, non-interacting model. The information obtained from mapping entropy and resolution, combined with the study of the correlation between neurons, allowed us to  approximately reconstruct the synaptic matrix of Hopfield models with $p\leq 4$; furthermore, by inspecting the properties of mappings in correspondence of decreases of the information loss rate we were able to ``detect'' the presence of biases in the patterns employed to build the synaptic matrix according to the Hebbian rule.

Moving to Hopfield models with $N=100$ spins, we showed that the minima of the mapping entropy feature a regular, linearly decreasing trend as a function of resolution, with the exception of a relatively flat region where the value of $S_{map}$ remains constant for a given number $n_{cg}$ of retained neurons. This behaviour allowed us to distinguish three regions and, correspondingly, three qualitatively distinct behaviours.

The first regime pertains to the most informative mappings with the lowest resolutions, or, equivalently, with the lowest $n_{cg}$. These mappings correspond to some of the most compressed representations that can be obtained at those values of $n_{cg}$ and their related semi-dispersion takes positive values. The conclusion is that, for low numbers of retained neurons, the mapping entropy selects those representations that describe specific clusters of neurons with strong internal interactions; the latter can thus be effectively decoupled from the rest of the network, which is treated as effective noise. This is the same simplified representation mechanism encountered for the $N=10$ model.

A rather opposite outcome is obtained for the second decimation regime, where we find the most informative mappings with higher resolutions and higher $n_{cg}$. These mappings correspond to some of the least compressed representations that can be obtained at those $n_{cg}$, and their related semi-dispersion values are low or even negative. For higher numbers of retained neurons, the mapping entropy optimisation workflow selects those representations that contain spins having weak couplings between each other, and strong couplings with the discarded ones; the optimal description of the entire network state is thus obtained retaining those that strongly correlate with the neglected ones.

The third regime, whereby the mapping entropy remains constant, is made up of equally informative mappings that, however, vary appreciably in terms of resolution as well as semi-dispersion. These representations entail the same amount of information in spite of the fact that their ``decimation strategy'' shifts continuously from the first to the second; such a degeneracy indicates the existence of an ``information plateau'', a buffer region where varying the representations of the system does not appreciably increases or decreases its informativeness.

The results presented in this work have highlighted nontrivial relations between the level of detail of the low-resolution representation at which the system is inspected and the underlying generative process, specifically the interaction matrix of the model. The analysis has focused on particular realisation of the Hopfield network, with the aim of establishing a link the most direct possible between a case-specific system and its optimal low-resolution mappings. The picture that emerges thus provides promising evidence that such link can be leveraged to extract nontrivial information about the properties of systems when only part of them is accessible to inspection.

In conclusion, this work has shown that it is possible to gather useful knowledge about a neural network through the analysis of the information content of its low-resolution representations; in particular, we have seen that the elements of the network that are pinpointed as the most informative ones depend on the specific decimated representation resolution level at which the system is described: this result can serve as a guide in the study of complex systems as well as in the construction of effective models of the latter, and paves the way to the development of a semi-automated protocol for the analysis of limited data gathered from systems composed by a large number of constituents.

\section*{Acknowledgments}

The authors are indebted with Matteo Marsili and Margherita Mele for an insightful reading of the manuscript and useful comments.
RP acknowledges support from ICSC - Centro Nazionale di Ricerca in HPC, Big Data and Quantum Computing, funded by the European Union under NextGenerationEU. Views and opinions expressed are however those of the author(s) only and do not necessarily reflect those of the European Union or The European Research Executive Agency. Neither the European Union nor the granting authority can be held responsible for them.
Funded by the European Union under NextGenerationEU. PRIN 2022 PNRR Prot. n. P2022MTB7E.

\section*{Data and software availability}

Raw data produced and analyzed in this work are freely available on the Zenodo repository \href{https://doi.org/10.5281/zenodo.13940980}{https://doi.org/10.5281/zenodo.13940980}.

\section*{Author contributions}

RP conceived the study. RP and RM proposed the method. RA carried out the simulations and the preliminary data analyses. All authors contributed to the analysis and interpretation of the data. All authors drafted the paper, reviewed the results, and approved the final version of the manuscript.

%%%%%%%%%%%%%%%%%%%%%%%%%%%%%%%%%%%%%%%%%%%%%%%%%%%%%%%%%

\bibliographystyle{ieeetr}
\nocite{*}
\bibliography{main}

%\printbibliography

\end{document}